\documentstyle[12pt,axodraw]{article}

\def\slashchar#1{\setbox0=\hbox{$#1$}           
   \dimen0=\wd0                                 
   \setbox1=\hbox{/} \dimen1=\wd1               
   \ifdim\dimen0>\dimen1                        
      \rlap{\hbox to \dimen0{\hfil/\hfil}}      
      #1                                        
   \else                                        
      \rlap{\hbox to \dimen1{\hfil$#1$\hfil}}   
      /                                         
   \fi}                                         %
\renewcommand{\theequation}{\arabic{section}.\arabic{equation}}
\newcommand{\newc}{\newcommand}
\newc{\gsim}{\lower.7ex\hbox{$\;\stackrel{\textstyle>}{\sim}\;$}}
\newc{\lsim}{\lower.7ex\hbox{$\;\stackrel{\textstyle<}{\sim}\;$}}
\newc{\oneloop}{h}
\newc{\Nindex}{n}
\newc{\Iindex}{n}
\newc{\Jindex}{p}
\newc{\Kindex}{q}
\newc{\Lindex}{r}
\newc{\ahat}{\hat{a}}
\newc{\bhat}{\hat{b}}
\newc{\Nc}{N_{c}}
\newc{\CG}{C_G}
\newc{\FFbS}{\overline{FF}S}
\newc{\FFbV}{\overline{FF}V}
\newc{\FFbe}{\overline{FF}\epsilon}
\newc{\FSS}{F_{SS}}
\newc{\FSSS}{F_{SSS}}
\newc{\FFFS}{F_{FFS}}
\newc{\FFFbS}{F_{\overline{FF}S}}
\newc{\FSSV}{F_{SSV}}
\newc{\FVS}{F_{VS}}
\newc{\FVVS}{F_{VVS}}
\newc{\FFFV}{F_{FFV}}
\newc{\FFFbV}{F_{\overline{FF}V}}
\newc{\FVV}{F_{VV}}
\newc{\FVVV}{F_{VVV}}
\newc{\fggV}{f_{ggV}}
\newc{\Fgauge}{F_{\rm gauge}}
\newc{\fSS}{f_{SS}}
\newc{\fSSS}{f_{SSS}}
\newc{\fFFS}{f_{FFS}}
\newc{\fFFbS}{f_{\overline{FF}S}}
\newc{\fSSV}{f_{SSV}}
\newc{\fVVS}{f_{VVS}}
\newc{\fVS}{f_{VS}}
\newc{\fFFV}{f_{FFV}}
\newc{\fFFbV}{f_{\overline{FF}V}}
\newc{\fVV}{f_{VV}}
\newc{\fVVV}{f_{VVV}}
\newc{\fgauge}{f_{\rm gauge}}
\newc{\feS}{f_{\epsilon S}}
\newc{\feeS}{f_{\epsilon\epsilon S}}
\newc{\fFFe}{f_{FF\epsilon}}
\newc{\fFFbe}{f_{\overline{FF}\epsilon}}
\newc{\feeV}{f_{\epsilon\epsilon V}}
\newc{\feV}{f_{\epsilon V}}
\newc{\fee}{f_{\epsilon\epsilon }}
\newc{\jhat}{J}
\newc{\ihat}{I}
\newc{\lnbar}{\overline{\rm ln}}
\def\beq{\begin{eqnarray}}
\def\eeq{\end{eqnarray}}
\def\bea{\begin{eqnarray}}
\def\eea{\end{eqnarray}}
\catcode`@=11
\long\def\@caption#1[#2]#3{\par\addcontentsline{\csname
  ext@#1\endcsname}{#1}{\protect\numberline{\csname
  the#1\endcsname}{\ignorespaces #2}}\begingroup
    \small
    \@parboxrestore
    \@makecaption{\csname fnum@#1\endcsname}{\ignorespaces #3}\par
  \endgroup}
\catcode`@=12

\begin{document}
\begin{titlepage}
\begin{flushright}
hep-ph/0111209 \\
FERMILAB-Pub-01/348-T\\
\end{flushright}
\vspace{0.3in}

\begin{center}
{\large\bf
Two-loop effective potential for a general renormalizable theory
and softly broken supersymmetry}
\end{center}
\vspace{.15in}

\begin{center}
{\sc Stephen P.~Martin}

\vspace{.1in}
{\it Department of Physics, Northern Illinois University, DeKalb IL 60115
{$\rm and$}\\
}{\it Fermi National Accelerator Laboratory,
P.O. Box 500, Batavia IL 60510 \\} 
\end{center}

\vspace{0.8in}

\begin{abstract}\noindent I compute the two-loop effective potential in
the Landau gauge for a general renormalizable field theory in four
dimensions. Results are presented for the $\overline{\rm MS}$
renormalization scheme based on dimensional regularization, and for the
$\overline{\rm DR}$ and $\overline{\rm DR}'$ schemes based on
regularization by dimensional reduction. The last of these is appropriate
for models with softly broken supersymmetry, such as the
Minimal Supersymmetric Standard Model. I find the parameter redefinition
which relates the $\overline{\rm DR}$ and $\overline{\rm DR}'$ schemes at
two-loop order. I also discuss the renormalization group invariance of the
two-loop effective potential, and compute the anomalous dimensions for
scalars and the beta function for the vacuum energy at two-loop order in
softly broken supersymmetry. Several illustrative examples and consistency
checks are included.

\end{abstract}
\end{titlepage}
\baselineskip=23pt
\baselineskip=20pt
\setcounter{footnote}{1}
\setcounter{page}{2}
\setcounter{figure}{0}
\setcounter{table}{0}

\tableofcontents

\section{Introduction}
\label{sec:intro}
\setcounter{equation}{0}
\setcounter{footnote}{1}

The Fermilab Tevatron collider and the CERN LHC collider hold the promise
of exposing the nature of spontaneous electroweak symmetry breaking. In
the Standard Model, this mechanism relies on a non-zero vacuum expectation
value (VEV) for a fundamental Higgs scalar field. There are good
theoretical and experimental reasons to suspect that this picture is
correct, but
incomplete, and must be embedded in a larger theory such as supersymmetry
\cite{WessBagger,Martin:1997ns}. When new experimental
discoveries are made, the tasks of 
telling the difference between different candidate models of electroweak
symmetry breaking and constraining the underlying parameters of the
successful theory will require high-precision calculational tools at the
two-loop level or better.

The effective potential 
\cite{Coleman:1973jx}-\cite{Sher:1989mj}
allows the calculation
of the VEVs in the true vacuum state of a theory with spontaneous symmetry
breaking.
In this formalism, the scalar fields of the theory are each separated into
a constant classical background $\phi$ plus quantum fluctuations. The
effective potential $V(\phi)$ is equal to the tree-level potential in the
classical background, plus the sum of one-particle-irreducible connected
vacuum graphs. These are calculated using the Feynman rules with
$\phi$-dependent masses and couplings. Thus one may write
\beq
V = V^{(0)}
+ {1\over 16 \pi^2} V^{(1)}
+ {1\over (16 \pi^2)^2} V^{(2)}
+ \ldots,
\eeq
where $V^{(n)}$ represents the $n$-loop correction.\footnote{To save ink, 
a factor of $1/(16 \pi^2)^n$ is always factored out of
the $n$-loop contribution to the loop expansion of the effective
potential, $\beta$-functions, and anomalous dimensions in this paper.}
In this paper, I will be concerned with the effective potential in
Landau gauge. Although the effective potential itself is
gauge-dependent, physical properties following from it, such as its
value at stationary points, and the question of whether or not
spontaneous symmetry breaking occurs, are gauge invariant 
\cite{gaugeindependence}.
The
one-loop
contribution $V^{(1)}$ is well-known for a general
field theory, and is reviewed in section \ref{sec:oneloop}. In
ref.~\cite{Ford:1992pn},
Ford, Jack and Jones have calculated $V^{(2)}$ in the special case of 
the
Standard Model using dimensional regularization (DREG) with
minimal subtraction or modified
minimal
subtraction ($\overline{\rm MS}$). Their calculations can be 
generalized to obtain the corresponding result for any renormalizable 
field
theory, as I
will do explicitly in section \ref{sec:msbar}.

However, it is well-known that the DREG regularization method is not
convenient for theories based on supersymmetry. This is because in DREG,
the vector field only has $4-2 \epsilon$ components, introducing a
spurious non-supersymmetric
mismatch with the number of degrees of freedom of the gaugino. Therefore,
in DREG the relationships between couplings which should hold in a
softly broken supersymmetric theory are violated even at one-loop order.
Instead, one can use the dimensional reduction (DRED) method \cite{DRED}, 
in which loop integrals are still
regularized by taking momenta in $4 - 2 \epsilon$ dimensions,
but all $4$ components of each vector field are kept. 
The extra
$2 \epsilon$ components of the gauge field in DRED transform like
scalars in the 
adjoint representation of the gauge group, and are known as epsilon
scalars. The renormalization scheme based on DRED with modified minimal
subtraction is known as $\overline{\rm DR}$. It has the virtue of
maintaining
manifest supersymmetry in theories where supersymmetry is not explicitly
broken.

Realistic models of the physics at the TeV scale do involve
explicit soft violations of supersymmetry, however. In such models, the
$\overline{\rm DR}$ renormalized dimensionless 
couplings of the theory obey the
relations prescribed by unbroken supersymmetry. However, the epsilon
scalars in general do not have the same masses or dimensionful couplings
as do the ordinary $4- 2 \epsilon$ vector field. In fact, computation of
the renormalization group (RG) equations shows that the running squared
masses
of the epsilon scalars cannot be consistently set equal to those of the
corresponding vector gauge bosons 
\cite{Jack:1994kd}.
This makes the $\overline{\rm DR}$ scheme also inconvenient, since the
epsilon-scalar masses are unphysical. A better scheme is
the $\overline{\rm DR}'$ scheme \cite{Jack:1994rk}, which differs from 
$\overline{\rm DR}$ by a parameter redefinition. The $\overline{\rm DR}'$
scheme offers the advantages that the epsilon-scalar masses completely
decouple from all RG equations, and also from the
equations that relate running renormalized parameters to pole masses and
other physical
observables. 

In this paper, I will present results for the two-loop effective potential
in the Landau gauge and in each of the $\overline{\rm MS}$, $\overline{\rm
DR}$, and $\overline{\rm DR}'$ renormalization schemes. For models with
exact supersymmetry, the last two schemes are the same, while for models
with softly broken supersymmetry the $\overline{\rm DR}'$ scheme is by far
the most convenient.

\begin{figure}[tb]
\SetWidth{1}
\hspace{0.5in}
\begin{picture}(112,136)(-56,-70)
\CArc(0,0)(36,0,180)
\CArc(0,0)(36,180,360)
\end{picture}
\hspace{2cm}
\begin{picture}(86,136)(-43,-70)
\CArc(0,25)(25,-90,270)
\CArc(0,-25)(25,90,450)
\Vertex(0,0){2}
\end{picture}
\begin{picture}(112,136)(-56,-70)
\CArc(0,0)(36,0,180)
\CArc(0,0)(36,180,360)
\Line(-36,0)(36,0)
\Vertex(-36,0){2}
\Vertex(36,0){2}
\end{picture}
\caption{Topologies of one-particle-irreducible connected vacuum Feynman
diagrams
for the one-loop and two-loop contributions to the effective potential.}
\label{fig:topologies}
\end{figure}
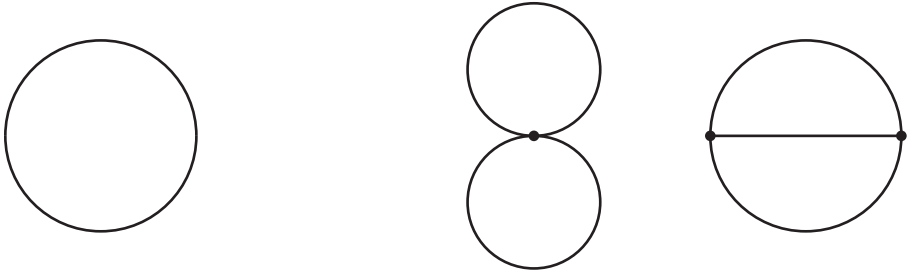

The topologies of the one-particle-irreducible connected vacuum graphs
at one- and two-loop orders are shown in Figure \ref{fig:topologies}.
Because the one-loop graph topology does not involve interaction
vertices, $V^{(1)}$ clearly depends only on the field-dependent
squared masses $m^2_{\Iindex}$, where the index $\Iindex$ runs over all
of the real scalars, two-component fermions, and vector degrees of
freedom in the theory. Note that any complex scalar can be written
in terms of two real scalars, while four-component Dirac and Majorana
fermions can always be written in terms of two-component left-handed Weyl
fermions, in a way throughly familiar to disciples of supersymmetry (see
refs.~\cite{WessBagger,Martin:1997ns} for a discussion). 
In any dimensional-continuation regularization scheme,
quadratic divergences are automatically discarded, and one finds
for the renormalized effective potential at one-loop order:
\beq
V^{(1)} = {1\over 4} \sum_{\Iindex} (-1)^{2 s_{\Iindex}} (2 s_{\Iindex} + 
1) (m_{\Iindex}^2)^2
(\lnbar m_{\Iindex}^2 - k_{\Iindex} )  .
\eeq
Here I have adopted the notation
\beq
\lnbar(x) = \ln(x/Q^2)  ,
\eeq
where $Q$ is the renormalization scale, and
$s_{\Iindex} = 0, 1/2, 1$ for real scalars, two-component fermion, and
vector
degrees of freedom 
respectively,\footnote{The contribution of epsilon
scalars is discussed in section \ref{sec:oneloop}.}
and $k_{\Iindex}$ are
constants depending on the details of the renormalization scheme. 

From figure \ref{fig:topologies}, it is clear that at two-loop order
the result must be of the form
\beq
V^{(2)} = 
\sum_{\Iindex,\Jindex} g^{\Iindex\Iindex\Jindex\Jindex} f_{\Iindex\Jindex} 
(m^2_\Iindex, m^2_\Jindex) +
\sum_{\Iindex,\Jindex,\Kindex} |g^{\Iindex\Jindex\Kindex}|^2 
f_{\Iindex\Jindex\Kindex} (m^2_{\Iindex}, m^2_{\Jindex}, m^2_{\Kindex})   ,
\eeq
where $g^{\Iindex\Jindex\Kindex\Lindex}$ and $g^{\Iindex\Jindex\Kindex}$ 
are field-dependent
four- and three-particle couplings, and $f_{\Iindex\Jindex}(x,y)$ and
$f_{\Iindex\Jindex\Kindex}(x,y,z)$ are $Q$-dependent functions
obtained by
performing the
appropriate two-loop integrations. So the task is
to identify these objects
for each combination of particle types that can contribute.

The rest of this paper is organized as follows. Section 2 describes
the field-dependent masses and couplings, lists the relevant Feynman
diagrams, and presents necessary conventions. Section 3 reviews the
one-loop effective potential, distinguishing between the 
$\overline{\rm MS}$,
$\overline{\rm DR}$, and
$\overline{\rm DR}'$ schemes. Sections 4-6 present the results for
the two-loop effective potential contribution in each of those schemes.
Section 6 also explicitly gives the redefinitions necessary to go from
$\overline{\rm DR}$ to
$\overline{\rm DR}'$. 
Section 7 discusses the RG invariance of the effective
potential
in the $\overline{\rm DR}'$ scheme, and derives some necessary results
for the scalar anomalous dimension and vacuum energy beta function
in softly broken supersymmetry. Section 8 contains some illustrative
examples and consistency checks.

\section{Conventions and setup}
\label{sec:conventions}
\setcounter{equation}{0}
\setcounter{footnote}{1}

\subsection{Field-dependent masses and couplings}
\label{subsec:setup}

Let us write the quantum fields of a general renormalizable field theory
as a set of real scalars
$R'_i$, two-component
Weyl fermions $\psi'_I$, and vector fields $A^{\prime\mu}_a$. 
Scalar flavor indices are $i,j,k,\ldots$; fermion flavor indices
are 
$I,J,K,\ldots$; and $a,b,c,\ldots$ run over the adjoint
representation of the gauge group. Space-time vector indices 
are written as Greek letters $\mu,\nu,\rho,\ldots$. I use a metric
with signature ($-$$+$$+$$+$), and the notations for fermions follow
\cite{WessBagger,Martin:1997ns}.
The
primes are used to
indicate that these fields are not
squared-mass eigenstates. The kinetic part of the lagrangian includes
\beq
- {\cal L} = {1\over 2} m^2_{ij} R'_i R'_j 
+ {1\over 2} (m^{IJ} \psi'_I \psi'_J + {\rm
c.c.}) +
{1\over 2} m^2_{ab} A^{\prime\mu}_a A'_{\mu b}   .
\label{undiagonalized}
\eeq
The symmetric fermion mass matrix $m^{IJ}$ yields a
fermion squared-mass matrix
\beq
m^2_{IJ} = m^*_{IK} m^{KJ}.
\eeq
Then $m^2_{ij}$ and $m^2_{ab}$ are real
symmetric matrices, and $m^2_{IJ}$ is a Hermitian matrix, and in general 
they all depend on the classical background scalar
fields.
In order to calculate the effective
potential, the first step is to rotate to squared-mass eigenstate
bases $R_i$, $\psi_I$, $A^\mu_a$. This can be done by using orthogonal
matrices 
$N^{(S)}$, $N^{(V)}$ for the scalar and vector degrees of freedom,
and a unitary matrix $N^{(F)}$ for the fermion degrees of freedom.
So, the rotations
\beq
R'_i &=& N_{ji}^{(S)} R_j 
  ,\\
\psi'_I &=& N_{JI}^{(F)*} \psi_J 
  ,\\
A^{\mu\prime}_a &=& N_{ba}^{(V)} A^\mu_b
  ,
\eeq
are chosen such that:
\beq
N^{(S)}_{ik} m^2_{kl} N^{(S)}_{jl} &=& \delta_{ij} m^2_i 
\label{NS}
  ,\\
N^{(F)}_{IK} m^2_{KL} N^{(F)*}_{JL} &=& \delta_{IJ} m^2_I 
  ,\\
N^{(V)}_{ac} m^2_{cd} N^{(V)}_{bd} &=& \delta_{ab} m^2_a .
\label{NV}
\eeq
Here 
$m^2_i$, $m^2_I$, and $m^2_a$ are respectively the
scalar, fermion, and vector squared-mass eigenvalues which will appear in
the effective potential.
It should be noted that in general $N^{(F)}$ diagonalizes the
fermion squared-mass matrix $m^2_{IJ}$,
but need not diagonalize the fermion mass matrix $m^{IJ}$.
All that is required is that 
\beq
M^{IJ} = N^{(F)*}_{IK} m^{KL} N^{(F)*}_{JL}
\eeq
has a
block diagonal form, with non-zero entries only between states with the
same squared-mass eigenvalue.
Indeed, it is quite often not particularly desirable 
for
$N^{(F)}$ to diagonalize the fermion mass matrix, for example
in the case of charged Dirac fermions,
with doubly-degenerate eigenvalues for $m^2_{I}$,
where $M^{IJ}$ is best left
off-diagonal
in $2\times 2$ blocks. The matrix
$M^{IJ}$ and its complex conjugate $M_{IJ}^*$
will appear as mass
insertions.
In practical applications, the diagonalizations just described
are easily performed numerically using a computer, and under favorable
circumstances (such as those studied in section \ref{sec:examples})
they can be done analytically. In either case, the problem
amounts to finding the
orthonormal eigenvectors of $m^2_{ij}$, $m^2_{IJ}$, and $m^2_{ab}$.
  
Now the interaction terms in a general renormalizable
theory can be written in terms of the squared-mass eigenstate fields
as
\beq
{\cal L}_{\rm S} &=& - {1\over 6} \lambda^{ijk} R_i R_j R_k
- {1\over 24} \lambda^{ijkl} R_i R_j R_k R_l
,  \label{LS}
\\
{\cal L}_{\rm SF} &=& - {1\over 2} y^{IJk} \psi_I \psi_J R_k + {\rm
c.c.}
, 
\\
{\cal L}_{\rm SV} &=& - {1\over 2} g^{abi} A_\mu^a A^{\mu b} R_i
-{1\over 4} g^{abij}  A_\mu^a A^{\mu b} R_i R_j
- g^{aij} A_\mu^a R_i \partial^\mu R_j 
, 
\label{LSV}
\\
{\cal L}_{\rm FV} &=& g^{aJ}_I A^a_\mu \psi^{\dagger I} {\overline
\sigma}^\mu \psi_J
,
\\
{\cal L}_{\rm gauge} &=& 
g^{abc} A^a_\mu A^b_\nu \partial^\mu A^{\nu c} 
- {1\over 4} g^{abe} g^{cde} A^{\mu a} A^{\nu b} A_\mu^c A_\nu^d
+ g^{abc} A^a_\mu \omega^b \partial^\mu \overline \omega^{c} ,
\label{Lgauge}
\eeq
where $\omega^a$ and $\overline{\omega}^a$ are massless (in Landau gauge) 
ghost 
fields. 
This defines the field-dependent couplings
to be used in the two-loop effective potential calculation.
The scalar interaction couplings $\lambda^{ijk}$ and $\lambda^{ijkl}$
are each completely symmetric under interchange of indices, and real.
The Yukawa couplings $y^{IJk}$ are symmetric under interchange of
the fermion flavor indices $I,J$. The remaining couplings all have
their
origins in gauge interactions.
The vector-scalar-scalar coupling $g^{aij}$ is
antisymmetric under interchange of $i,j$. The pure gauge interaction
$g^{abc}$ is completely antisymmetric; it is determined by the 
original gauge
coupling $g$, the antisymmetric
structure constants $f^{abc}$ of the gauge group, and $N^{(V)}$, according 
to
\beq
g^{abc} = g f^{efg} 
N^{(V)}_{ae}
N^{(V)}_{bf}
N^{(V)}_{cg}.
\label{definegabc}
\eeq
Similarly, if the fermions $\psi'_I$ transform under the gauge group with
representation matrices $(T^a)_I^J$, then the vector-fermion-fermion
couplings are
\beq
g^{aJ}_I = g (T^b)^K_L N^{(F)*}_{JK} N^{(F)}_{IL} N^{(V)}_{ab} .
\label{definegapq}
\eeq
Note that even the dimensionless couplings generically depend on the
classical
scalar background fields $\phi$, through their dependence on the
rotation matrices $N^{(S)}$, $N^{(F)}$, and $N^{(V)}$.

\subsection{The Feynman diagrams}

\begin{figure}[tbp]
\SetWidth{1}
\begin{picture}(112,136)(-56,-70)
\DashCArc(0,0)(36,0,180){6}
\DashCArc(0,0)(36,180,360){6}
\DashLine(-36,0)(36,0){6}
\Vertex(-36,0){2}
\Vertex(36,0){2}
\Text(0,-62)[]{$SSS$}
\end{picture}
\begin{picture}(86,136)(-43,-70)
\DashCArc(0,25)(25,-90,270){6}
\DashCArc(0,-25)(25,90,450){6}
\Vertex(0,0){2}
\Text(0,-62)[]{$SS$}
\end{picture}
\begin{picture}(112,136)(-56,-70)
\ArrowArcn(0,0)(36,180,0)
\ArrowArc(0,0)(36,180,360)
\DashLine(-36,0)(36,0){6}
\Vertex(-36,0){2}
\Vertex(36,0){2}
\Text(0,-62)[]{$FFS$}
\end{picture}
\begin{picture}(112,136)(-56,-70)
\ArrowArcn(0,0)(36,180,90)
\ArrowArc(0,0)(36,-180,-90)
\ArrowArc(0,0)(36,0,90)
\ArrowArcn(0,0)(36,0,-90)
\DashLine(-36,0)(36,0){6}
\Vertex(-36,0){2}
\Vertex(36,0){2}
\Vertex(0,-36){3}
\Vertex(0,36){3}
\Text(0,-62)[]{$\overline{FF}S$}
\end{picture}

\noindent
\begin{picture}(112,136)(-56,-70)
\DashCArc(0,0)(36,0,180){6}
\DashCArc(0,0)(36,180,360){6}
\Photon(-36,0)(36,0){2.2}{4.5}
\Vertex(-36,0){2}
\Vertex(36,0){2}
\Text(0,-62)[]{$SSV$}
\end{picture}
\begin{picture}(86,136)(-43,-70)
\DashCArc(0,25)(25,-90,270){6}
\PhotonArc(0,-25)(25,90,450){-2.2}{8.5}
\Vertex(0,0){2}
\Text(0,-62)[]{$VS$}
\end{picture}
\begin{picture}(112,136)(-56,-70)
\PhotonArc(0,0)(36,0,180){2.2}{6.5}
\PhotonArc(0,0)(36,180,360){2.2}{6.5}
\DashLine(-36,0)(36,0){6}
\Vertex(-36,0){2}
\Vertex(36,0){2}
\Text(0,-62)[]{$VVS$}
\end{picture}
\begin{picture}(112,136)(-56,-70)
\ArrowArc(0,0)(36,0,180)
\ArrowArc(0,0)(36,180,360)
\Photon(-36,0)(36,0){2.2}{4.5}
\Vertex(-36,0){2}
\Vertex(36,0){2}
\Text(0,-62)[]{$FFV$}
\end{picture}

\noindent
\begin{picture}(112,136)(-56,-70)
\ArrowArcn(0,0)(36,180,90)
\ArrowArc(0,0)(36,0,90)
\ArrowArcn(0,0)(36,270,180)
\ArrowArc(0,0)(36,-90,0)
\Photon(-36,0)(36,0){2.2}{4.5}
\Vertex(-36,0){2}
\Vertex(36,0){2}
\Vertex(0,-36){3}
\Vertex(0,36){3}
\Text(0,-62)[]{$\overline{FF}V$}
\end{picture}
\begin{picture}(86,136)(-43,-70)
\PhotonArc(0,25)(25,-90,270){-2.2}{8.5}
\PhotonArc(0,-25)(25,90,450){-2.2}{8.5}
\Vertex(0,0){2}
\Text(0,-62)[]{$VV$}
\end{picture}
\begin{picture}(112,136)(-56,-70)
\PhotonArc(0,0)(36,0,180){2.2}{6.5}
\PhotonArc(0,0)(36,180,360){2.2}{6.5}
\Photon(-36,0)(36,0){2.2}{4.5}
\Vertex(-36,0){2}
\Vertex(36,0){2}
\Text(0,-62)[]{$VVV$}
\end{picture}
\begin{picture}(112,136)(-56,-70)
\DashCArc(0,0)(36,0,180){3}
\DashCArc(0,0)(36,180,360){3}
\Photon(-36,0)(36,0){2.2}{4.5}
\Vertex(-36,0){2}
\Vertex(36,0){2}
\Text(0,-62)[]{$ggV$}
\end{picture}
\caption{The one-particle-irreducible connected Feynman diagrams
contributing to the two-loop effective potential. Dashed lines
denote real scalars, solid lines denote Weyl fermions carrying
helicity along the arrow direction, wavy lines are for vector
bosons, and dotted lines are for ghosts. The large dots between
opposing arrows on the
fermion lines in the $\FFbS$ and $\FFbV$
diagrams denote mass insertions. The $\FFbS$ diagram
is accompanied by its complex conjugate (the same diagram with all arrows
reversed).}
\label{fig:feynmandiagrams}
\end{figure}
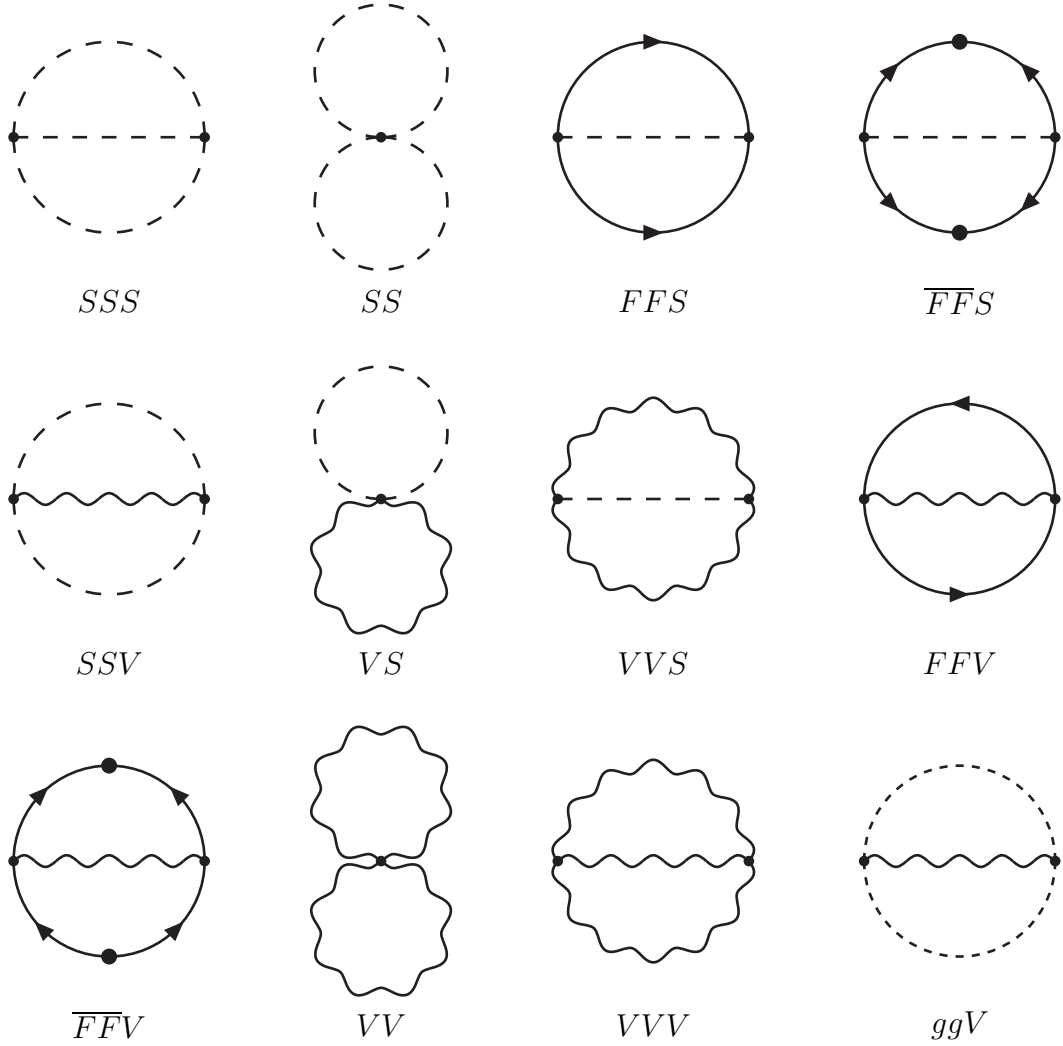

The two-loop effective potential is to be evaluated by
computing the one-particle-irreducible connected vacuum Feynman diagrams
appearing in figure \ref{fig:feynmandiagrams}. 
The masses and couplings of fields appearing in these diagrams are as
indicated on the right sides of eqs.~(\ref{NS})-(\ref{Lgauge}). Dashed
lines denote scalar propagators. Solid lines denote fermion
propagators with helicity along the direction of the arrow, and large dots
between opposing arrows denote insertions of the fermion mass matrix
$M^{IJ}$ or its complex conjugate $M^*_{IJ}$, depending on whether the
arrows are incoming or outgoing. Vector propagators are indicated by wavy
lines, and ghost propagators by dotted lines. Each graph is also labelled
by the type of propagators it contains, with $S,F,V,g$ standing
respectively for scalar, fermion, vector and ghost. Also, the presence of
mass insertions in fermion lines is indicated by the overlines in the
labels $\FFbS$ and $\FFbV$. The results for these Feynman
diagrams (plus counterterms) are
reported in sections \ref{sec:msbar}, \ref{sec:drbar}, and
\ref{sec:drbarprime}.

\subsection{Two-loop integral functions needed for vacuum graphs}

All of the effective potential two-loop integrals can be expressed in
terms of linear combinations of functions 
introduced and studied by Ford, Jack and
Jones in \cite{Ford:1992pn}. 
I will follow a notation similar but not identical to
theirs:
the functions $\ihat(x,y,z)$, $\jhat (x,y)$, and $J(x)$ used here
are equal to the $\epsilon$-independent parts of the functions $\hat
I (x,y,z)$, $\hat J(x,y)$,
and $J(x)$ used in ref.~\cite{Ford:1992pn}, up to obvious factors of
$1/16 \pi^2$. Explicitly,
I choose to express results in terms of:
\beq
J(x) &=& 
x (\lnbar x -1)
, \\
\jhat (x,y) &=& 
x y (\lnbar x -1)(\lnbar y -1)
, \\
\ihat(x,y,z) &=& 
{1\over 2} (x-y-z) \lnbar y \lnbar z
+ {1\over 2}(y-x-z) \lnbar x \lnbar z
+ {1\over 2}(z-x-y) \lnbar x \lnbar y
\nonumber \\ && 
+ 2 x \lnbar x + 2 y \lnbar y + 2 z \lnbar z 
-{5\over 2} (x+y+z)
- {1\over 2} \xi (x,y,z) \> .
\eeq
Here $\xi(x,y,z)$ was originally found in terms of Lobachevskiy's function
or related integrals in ref.~\cite{Ford:1992pn} using methods developed
in 
\cite{Kotikov,Ford:1992hw}, but it can also be expressed 
\cite{Davydychev:1993mt,Caffo:1998du,Espinosa:2000df}
in terms of
dilogarithms according to (for $x,y \leq z$):
\beq
\xi (x,y,z) &=& R \Bigl \lbrace
2 
{\rm ln} [(z+x-y - R)/2z ]
{\rm ln} [(z+y-x - R)/2z ] - {\rm ln}(x/z) {\rm ln}(y/z)
\nonumber \\ && 
- 2 {\rm Li}_2[(z+x-y - R)/2z ]- 2{\rm Li}_2[(z+y-x - R)/2z ]
+ \pi^2/3 \Bigr \rbrace
\label{definexi}
\eeq
with 
\beq
R = [x^2 + y^2 + z^2 - 2 x y - 2 x z - 2 y z]^{1/2} .
\eeq
The dilogarithm function is defined in the standard \cite{Lewin} way
for any complex $z$:
\beq
{\rm Li}_2 (z) = -\int_0^z {{\rm ln}(1-t) \over t} dt 
  .
\eeq
To resolve branch cut ambiguities which could arise,
consistently choose
$
{\rm Arg}(R) = 0\>\,{\rm or}\>\, \pi/2
$
along with
\beq
-\pi < {\rm Im}[ {\rm ln}(z)] \leq \pi
\eeq
for all logarithms of negative or complex $z$, including the logarithm
appearing in
the integral
definition of
the dilogarithm. So, for example, when $x$ is real and greater than 1,
\beq
{\rm Im}[ {\rm Li}_2(x)] &=& - i \pi {\rm ln}(x) ,\\
{\rm Im}[ {\rm Li}_2(x \pm i \delta)] &=& \pm i \pi {\rm ln}(x) ,
\eeq
for $\delta$ real and infinitesimal,
while ${\rm Im}[ {\rm Li}_2(x)]=0$ if $x$ is real and less than 1.
The functions $\xi (x,y,z)$  and therefore $\ihat (x,y,z)$ are invariant
under interchange of any two of $x,y,z$.

It is useful to have expressions for these functions in the special cases
of vanishing arguments.
In addition to the trivial identities $J(0) = 0$, $\jhat (x,0) = \jhat
(0,x) = 0$ and $\ihat (0,0,0)=0$, one finds
\cite{Espinosa:2000df}:
\beq
\ihat (0,x,y) &=& 
(x-y) \left [ {\rm Li}_2(y/x) - {\rm \ln}(x/y) \lnbar
(x-y) + {1\over 2} (\lnbar x)^2 - {\pi^2 \over 6} \right ]
\nonumber \\  && 
-{5\over 2} (x+y)
+ 2 x \lnbar x + 2 y \lnbar y
- x \lnbar x \lnbar y
  ,\\
\ihat (0,x,x) &=& 
2 J(x) - 2 x - {1\over x} \jhat (x,x)
\, =\, -x (\lnbar x)^2 + 4 x \lnbar x -5 x ,
\label{I0xx}
\\
\ihat (0,0,x) &=& -{1\over 2} x (\lnbar x)^2 + 2 x \lnbar x - {5\over 2}x
- {\pi^2 \over 6}x 
  .
\eeq
It is also sometimes useful to expand these functions for infinitesimal
arguments:
\beq
\ihat (\delta,x,y) &=& 
\ihat (0,x,y) 
+ \delta \Bigl \lbrace
- (x+y) \ihat (0,x,y) 
- 2 \jhat (x,y) + 3 x J(x) + 3 y J(y)
- y J(x) 
\nonumber \\ &&
- x J(y) - (x+y)^2 + (x-y)[J(y) - J(x)]\lnbar\delta
\Bigr \rbrace /(x-y)^2 + \ldots
  ,
\label{deltaforce}
\\
\ihat (\delta,x,x) &=& 2 J(x) - 2 x - {1\over x} \jhat (x,x)
\nonumber \\ &&
+\delta \left  \lbrace 4 + {1\over 2 x^2} \jhat (x,x) + {3\over x} J(x) -
\bigl [1+ {J(x)\over x} \bigr ]\lnbar \delta \right \rbrace 
+ \ldots  ,
\\
\ihat (\delta_1, \delta_2, x) &=&
\ihat (0,0,x) 
+ {\delta_1 \over x} \left [
-x - \ihat (0,0,x) + 3 J(x) - J(x) \lnbar\delta_1 \right ]
\nonumber \\ && 
+ {\delta_2 \over x} \left [
-x - \ihat (0,0,x) + 3 J(x) - J(x) \lnbar\delta_2 \right ]
+ {\delta_1 \delta_2 \over x^2}
\Bigl \lbrace - 2 \ihat (0,0,x) 
\nonumber \\ && 
+ 4 J(x) 
- 2 x - 
[x + J(x)] (\lnbar\delta_1 + \lnbar\delta_2) 
+ x \lnbar\delta_1 \lnbar\delta_2 \Bigr \rbrace
+ \ldots   ,
\label{sealteamsix}
\eeq
where the ellipses stand for terms with more than one power of $\delta$ or
either $\delta_1$ or $\delta_2$.

\subsection{Conventions for softly-broken supersymmetric models}
\label{subsec:soft}

One of the main applications of the results of this paper
is to models with softly broken supersymmetry, such as the Minimal
Supersymmetric Standard Model (MSSM). Therefore I now list the
relevant conventions to be used here for such models.
In general, the superpotential is given in terms of the chiral
superfields $\Phi_i$ by
\beq
W = {1\over 6} Y^{ijk} \Phi_i \Phi_j \Phi_k +
{1\over 2} \mu^{ij} \Phi_i \Phi_j ,
\eeq
and the soft supersymmetry-breaking part of the Lagrangian is
\beq
-{\cal L}_{\rm soft} = 
({1\over 6} a^{ijk} \phi_i \phi_j \phi_k 
+{1\over 2} b^{ij} \phi_i \phi_j +c^i \phi_i + {1\over 2} M
\lambda_a \lambda_a + {\rm
c.c.}) +
(m^2)_i^j \phi^{*i} \phi_j + \Lambda    ,
\label{softterms}
\eeq
where the $\phi_i$ are the complex scalar field components of the
$\Phi_i$, and the $\lambda_a$ are the two-component gaugino fermions
with mass $M$. 
The parameter $c^i$ can only appear if there is a gauge-singlet chiral
superfield in the theory. 
Note the presence of a vacuum energy term $\Lambda$. This is required in
order for the full effective potential to be RG
invariant 
\cite{Einhorn:1983pp}-%
\cite{Ford:1993mv}.
The
two-loop beta function for
$\Lambda$
is obtained in section \ref{sec:rginv}, and
the beta functions for each of the other couplings at two-loop
order are given in 
\cite{Martin:1994zk,Yamada:1994id,Jack:1994kd,Jack:1994rk}, 
Flipping the heights on
all indices of a coupling implies complex conjugation, so
$Y_{ijk} = (Y^{ijk})^*$,
$\mu_{ij} = (\mu^{ij})^*$,
$a_{ijk} = (a^{ijk})^*$, etc.
The representation matrices
for
the chiral superfields are denoted by $({T^a})_i^j$. They satisfy
\beq
[{T^a}, {T^b}] = i f^{abc} T^c   ,
\eeq
where $f^{abc}$ are the totally antisymmetric structure constants of
the gauge group $G$. Then
\beq
({T^a T^a})_i^j &=& C(i) \delta_i^j  ,\\
{\rm Tr}[{T^a T^b}] &=& S(R) \delta^{ab}  ,\\
f^{acd} f^{bcd} &=& \CG \delta^{ab}
\eeq
define the quadratic Casimir invariant $C(i)$ for each representation,
the total Dynkin index $S(R)$ summed over all representations, and
the Casimir invariant of the adjoint representation. The dimension of
the adjoint representation is
\beq
d_G = {\rm Tr}[C(i)]/S(R) .
\eeq
I use a normalization
such
that each fundamental representation of $SU(N)$ has $C(i) = (N^2-1)/2 N$
and contributes 1/2 to $S(R)$.

\section{One-loop effective potential in the 
$\overline{\rm MS}$,
$\overline{\rm DR}$, and 
$\overline{\rm DR}'$ schemes}
\label{sec:oneloop}
\setcounter{equation}{0}
\setcounter{footnote}{1}

In this section, I review the known answers for the one-loop
effective potential. This will allow us to carefully distinguish
the results in the $\overline{\rm MS}$, $\overline{\rm DR}$,
and $\overline{\rm DR}'$ schemes.

In the $\overline{\rm MS}$ scheme and Landau gauge, one has 
\beq
V^{(1)}_{\overline{\rm MS}} = 
V_S^{(1)} + V_F^{(1)} + V^{(1)}_V
\eeq
where the different contributions arise from scalars, fermions,
and vectors going around the loop in figure \ref{fig:topologies}:
\beq
V_S^{(1)}  &=& {1\over 4} \sum_i (m_i^2)^2 (\lnbar m_i^2 - 3/2)  ,\\
V_F^{(1)}  &=& -{1\over 2} \sum_I (m_I^2)^2 (\lnbar m_I^2 - 3/2)  ,\\
V_V^{(1)}  &=& {3\over 4} \sum_a (m_a^2)^2 (\lnbar m_a^2 - 5/6)   .
\eeq
The appearance of $5/6$ rather than $3/2$ in $V_V^{(1)}$ is due
to the fact that there are only $4 - 2 \epsilon$, rather than 4,
vector degrees of freedom in $\overline{\rm MS}$.

In the $\overline{\rm DR}$ scheme, one must include also the effects of
the epsilon scalars. Now, it is tempting to assume that the epsilon
scalars have the same field-dependent mass as their $4- 2\epsilon$ vector
counterparts. However, as pointed out in ref.~\cite{Jack:1994kd}, this is
actually
inconsistent except in models with exact supersymmetry, unless one sticks
to only one fixed value of the renormalization scale $Q$, because the
epsilon-scalar squared mass has a beta function which is not homogeneous.
Therefore, in general one must allow the epsilon scalars to have 
squared-mass eigenvalues $\hat m^2_{a}$ which are distinct from the
$m^2_a$
for the ordinary vectors. 
To be specific, consider the explicit 
form of the field-dependent squared-mass matrix for the ordinary $4 - 2
\epsilon$ vector fields:
\beq
m^2_{ab} = g_a g_b \phi^{* i}\{ T^a ,T^b \}_i^j \phi_j
  .
\label{coolvectormasses}
\eeq
This has eigenvalues $m^2_a$.
For the epsilon-scalar squared-mass matrix, one has instead:
\beq
\hat m^2_{ab} = 
m^2_{ab} + \delta_{ab} m_\epsilon^2
  ,
\label{escalarmasses}
\eeq
where $m_{\epsilon}^2$ is an ``evanescent" \cite{evanescent} parameter.
This matrix requires an orthogonal
diagonalization matrix $N^{(\epsilon)}$ which differs from $N^{(V)}$:
\beq
N^{(\epsilon)}_{a c} 
\hat m^2_{cd}
N^{(\epsilon)}_{b d}
= \delta_{a b} \hat m^2_{a}  .
\eeq 
Unless supersymmetry is not explicitly broken, 
the eigenvalues $\hat m^2_{a}$ will 
in general 
differ
from $m^2_a$, and the corresponding couplings of the squared-mass
eigenstate epsilon scalars
are different from the couplings of squared-mass eigenstate vectors,
because $N^{(\epsilon)}$
differs from $N^{(V)}$. 

In the $\overline{\rm DR}$ scheme, with epsilon scalars included, one now 
finds
\beq
V^{(1)}_{\overline{\rm DR}} = V_S^{(1)} + V_F^{(1)} + V^{(1)}_V
+ V_\epsilon^{(1)}
\eeq
where $V_S^{(1)}$, $V_F^{(1)}$, $V^{(1)}_V$ are as before, and
\beq
V_\epsilon^{(1)}  &=& -{1\over 2}\sum_{a} (\hat m_{a}^2)^2 .
\eeq

However, $m^2_\epsilon$ is an additional parameter with 
no physically
observable counterpart, and so its appearance in the effective potential
is quite
inconvenient. The functional form of the effective potential is also not
directly physically
observable, so there is no contradiction; $m^2_\epsilon$
must cancel only from observable quantities. However, clearly one would
like to avoid having to include a distinct epsilon-scalar mass in
calculations in the first place.
This
problem was solved in the context
of softly broken supersymmetric models in ref.~\cite{Jack:1994rk} with the
introduction of the $\overline{\rm DR}'$ scheme. 
The point is that one can remove the dependence of the full
one-loop effective
potential on
$m_\epsilon^2$ by redefining the ordinary scalar squared
masses and the vacuum energy term appearing in the
tree-level part eq.~(\ref{softterms}):
\beq
(m^2_{\overline{\rm DR}'})_i^j &=& (m^2_{\overline{\rm DR}})_i^j 
-{1\over 16 \pi^2} \left [ \delta_i^j 2 g^2 C(i)  
m_{\epsilon}^2
\right ]   ,
\label{tenthmountain}
\\
\Lambda_{\overline{\rm DR}'} &=& \Lambda_{\overline{\rm DR}}
- {1\over 16 \pi^2} {d_G (m_\epsilon^2)^2\over 2}
  .
\eeq
(These hold for a simple or $U(1)$ gauge group. If there is more
than one simple or $U(1)$ gauge group, then the correction
terms should be summed over subgroups, with a distinct $m_\epsilon^2$
for each subgroup.)
The result is the $\overline{\rm DR}'$ scheme,
and the effective potential in this scheme is the one usually
quoted in the literature (and often slightly incorrectly referred to
as the $\overline{\rm DR}$ one):
\beq
V^{(1)}_{\overline{\rm DR}'} &=& 
\sum_{\Nindex} (-1)^{2 s_{\Nindex}} (2 s_{\Nindex} + 1) \oneloop 
(m_{\Nindex}^2) 
=  {\rm STr}\left [ \oneloop (m_{\Nindex}^2) \right ] ,
\label{drbarprimeoneloop}
\eeq
where 
\beq
\oneloop (x) = {x^2\over 4} \left [ \lnbar (x) - 3/2 \right ], 
\label{defineoneloop}
\eeq
and $\Nindex$ runs over all real scalar, Weyl fermion, and vector
degrees of freedom. The scalar squared masses occurring
in eq.~(\ref{drbarprimeoneloop}) are the ones following from the
redefinition in eq.~(\ref{tenthmountain}), and the vector squared masses
are the eigenvalues of eq.~(\ref{coolvectormasses}). The 
$\overline{\rm DR}'$ effective
potential is both manifestly supersymmetric when the soft terms vanish, 
and
independent of the unphysical evanescent parameter $m^2_{\epsilon}$ when
the soft terms do not vanish. It
is not hard to see
that
$m^2_{\epsilon}$ is simultaneously banished from the equations which
relate the
physical pole masses to the $Q$-dependent running masses in the theory,
so it has been successfully decoupled from all practical calculations.
It would be quite clumsy to use 
the original $\overline{\rm DR}$ scheme in studies of realistic models
like the
MSSM, since in RG running and evaluation of the
effective potential one would have to keep extra contributions from
epsilon-scalar masses in order to avoid inconsistencies. Therefore the
$\overline{\rm DR}'$ scheme is the preferred one.

After making this painful distinction, it must be admitted that the
$\overline{\rm DR}'$ final result for the effective potential has exactly
the same form that one would have obtained if one had naively set
$m^2_\epsilon$ equal to zero in the first place in the $\overline{\rm DR}$
scheme calculation. However, this naive procedure is technically
inconsistent whenever RG running is involved 
\cite{Jack:1994kd}
and does not work for 
other
calculations involving epsilon scalars, so one should really distinguish
between the two schemes as a matter of principle. The parameters
appearing in the $\overline{\rm DR}'$ effective potential obey
$\overline{\rm DR}'$ renormalization group equations, which 
differ from the
$\overline{\rm DR}$ ones with $m^2_\epsilon$ set equal to 0.

The procedure of going from the $\overline{\rm DR}$ scheme to the
$\overline{\rm DR}'$ scheme is similar at two loops, and is described
explicitly in 
section \ref{sec:drbarprime}.

\section{Two-loop effective potential in the 
$\overline{\rm MS}$ scheme}
\label{sec:msbar}
\setcounter{equation}{0}
\setcounter{footnote}{1}

The two-loop effective potential in the $\overline{\rm MS}$
scheme for the general theory with masses and couplings described by
eqs.~(\ref{NS})-(\ref{Lgauge})
can be computed by the methods described in \cite{Ford:1992pn}. In fact,
all of the hard work of evaluating the relevant Feynman loop integrals
has already been
accomplished there; no new types of integrals arise. Momentum integrals
and vector indices each run over $4 - 2 \epsilon$ dimensions.
For each two-loop
diagram, one must include counterterms for the various one-loop divergent
subdiagrams. The result still includes single and double poles in
$\epsilon$, which are then simply removed by two-loop counterterms in
modified minimal
subtraction. The final result can be divided into parts corresponding
to the various graphs of figure \ref{fig:feynmandiagrams}.
Because the $VV$, $VVV$, and $ggV$ graphs all involve the same
field-dependent coupling $g^{abc}$, it is natural to combine their
contributions into a pure gauge piece $V^{(2)}_{\rm gauge}$. 

For the result, I find:
\beq
V^{(2)} &=& 
V^{(2)}_{SSS}
+ V^{(2)}_{SS}
+ V^{(2)}_{FFS}
+ V^{(2)}_{\FFbS}
+ V^{(2)}_{SSV}
+ V^{(2)}_{VS}
+ V^{(2)}_{VVS}
\nonumber \\ &&
+ V^{(2)}_{FFV}
+ V^{(2)}_{\FFbV}
+ V^{(2)}_{\rm gauge}
  ,
\eeq
where, in terms of the masses and couplings as specified in
eqs.~(\ref{NS})-(\ref{Lgauge}),
\beq
V^{(2)}_{SSS} &=& {1\over 12} (\lambda^{ijk})^2 
                  \fSSS (m^2_i, m^2_j, m^2_k) 
  ,
\label{VSSSMSbar}
\\
V^{(2)}_{SS} &=& {1\over 8} \lambda^{iijj} \fSS (m^2_i, m^2_j) 
  ,\\
V^{(2)}_{FFS} &=& {1\over 2} |y^{IJk}|^2 \fFFS (m^2_I, m^2_J, m^2_k)
  ,\\
V^{(2)}_{\FFbS} &=& {1\over 4} y^{IJk} y^{I'J'k} M^*_{II'} M^*_{JJ'}
                             \fFFbS (m^2_I, m^2_J, m^2_k) + {\rm c.c.}
  ,\\
V^{(2)}_{SSV} &=& {1\over 4} (g^{aij})^2 \fSSV (m^2_i,m^2_j, m_a^2)
  ,\\
V^{(2)}_{VS} &=& {1\over 4} g^{aaii} \fVS (m^2_a, m^2_i) 
  ,\\
V^{(2)}_{VVS} &=& {1\over 4} (g^{abi})^2 \fVVS (m^2_a, m^2_b,m^2_i) 
  ,\\
V^{(2)}_{FFV} &=& {1\over 2} |g^{aJ}_I|^2 \fFFV (m^2_I, m^2_J, m^2_a)
  ,\\
V^{(2)}_{\FFbV} &=& {1\over 2} g^{aJ}_I g^{aJ'}_{I'} 
M^{II'} M^*_{JJ'} \fFFbV (m^2_I, m^2_J, m^2_a)
  ,\\
V^{(2)}_{\rm gauge} &=& {1\over 12} (g^{abc})^2 \fgauge 
(m^2_a,m^2_b,m^2_c)
  ,
\label{VgaugeMSbar}
\eeq
in which all indices on the right side are summed over.
The loop-integral functions appearing here are given by:
\beq
\fSSS (x,y,z) &=& -\ihat (x,y,z) 
  , \label{fSSS}
\\
\fSS (x,y) &=& \jhat (x,y)
  ,\label{fSS}
\\
\fFFS (x,y,z) &=& 
\jhat (x,y) -\jhat (x,z) - \jhat (y,z) + (x+y-z) \ihat (x,y,z) 
  ,
\label{fFFS}
\\
\fFFbS (x,y,z) &=& 2 \ihat (x,y,z) 
  ,
\label{fFFbS}
\\
\fSSV (x,y,z) &=&  {1\over z} \Bigl \lbrace 
   (-x^2-y^2-z^2 +2 x y +2 x z +2 y z) \ihat (x,y,z)  
   + (x-y)^2 \ihat (0,x,y) 
   \nonumber \\ && 
+ (y-x-z) \jhat (x,z) +(x-y-z) \jhat (y,z) 
+ z \jhat (x,y) 
\Bigr \rbrace
   \nonumber \\ &&
+ 2 (x + y -z/3) J(z)
  ,
\label{fSSV}
\\
\fVS (x,y) &=& 3 \jhat (x,y) + 2 x J(y)
  ,
\\
\fVVS (x,y,z) &=& {1 \over 4 x y}  \Bigl \lbrace
       (-x^2 - y^2 - z^2 - 10 x y +2 x z +2 y z) \ihat (x,y,z) 
\nonumber \\ && 
       +(x-z)^2 \ihat (0,x,z) +(y -  z)^2 \ihat (0,y,z) 
       - z^2 \ihat (0,0,z) 
\nonumber \\ && 
+(z-x-y) \jhat (x,y) +y \jhat(x,z) +x \jhat (y,z) 
\Bigr \rbrace
\nonumber \\ && 
+ {1\over 2} J(x) + {1\over 2} J(y) +2 J(z) 
- x-y-z
  ,
\\
\fFFV (x,y,z) &=& 
{1\over z} \Bigl \lbrace (x^2 +y^2 -2 z^2 -2 x y +x z+y z) \ihat (x,y,z) 
- (x-y)^2 \ihat (0,x,y) 
\nonumber \\ && 
+(x - y - 2 z) \jhat (x,z) + (y-x-2z) \jhat (y,z) 
+2 z \jhat (x,y) 
\Bigr\rbrace 
\nonumber \\ &&                 
+ 2 (-x - y +z/3) J(z) - 2 x J(x) - 2 y J(y)
+ (x+y)^2 -z^2
  , \\
\fFFbV (x,y,z) &=& 6 \ihat(x,y,z) + 2(x+ y+ z) -4 J(x) - 4 J(y)
  ,\\
\fgauge (x,y,z) &=& {1\over 4 x y z} \Bigl \lbrace
(-x^4 - 8 x^3 y - 8 x^3 z +32 x^2 y z + 18 y^2 z^2) \ihat (x,y,z)
\nonumber \\ &&
+ (y-z)^2 (y^2 + 10 y z + z^2) \ihat (0,y,z) 
+ x^2 (2 y z - x^2) \ihat (0,0,x)
\nonumber \\ && 
+ (x^2 - 9 y^2 - 9 z^2  +9 x y + 9 x z + 14 y z) x \jhat(y,z)
\nonumber \\ && 
+ 4 x^3 y z +48 x y^2 z^2 
+ (22 y + 22 z - 16x/3) x y z J(x)
\Bigr \rbrace 
\nonumber \\ &&
+ (x \leftrightarrow y) + (x \leftrightarrow z).
\label{fgaugeMSbar}
\eeq
Symmetry factors have been explicitly factored
out of eqs.~(\ref{VSSSMSbar})-(\ref{VgaugeMSbar}), but fermion-loop minus
signs and other factors associated with
the evaluation of the Feynman diagrams are contained in the
definitions of the functions. The functions obey obvious symmetries:
$\fSSS (x,y,z)$ and $\fgauge (x,y,z)$ are invariant under
interchange of any two of $x,y,z$, while $\fSS(x,y)$, $\fFFS (x,y,z)$,
$\fFFbS (x,y,z)$, $\fSSV (x,y,z)$, $\fVVS (x,y,z)$, $\fFFV (x,y,z)$,
and $\fFFbV (x,y,z)$ are each invariant under interchange of $x,y$.

The functions involving vector fields contain factors $1/x$, $1/y$,
and $1/z$ which
appear
to be singular in the massless vector limit. This is due to the appearance
in the Landau gauge of vector propagators
\beq
{1\over i} \left ( {\eta^{\mu\nu} - p^\mu p^\nu/p^2 \over
p^2 + m^2 - i \epsilon } \right )
  ,
\eeq
which give rise to factors 
\beq
{1\over p^2 (p^2 + m^2)} 
=
{1\over m^2}\left (
{1\over p^2} - {1\over p^2 + m^2} \right )
\eeq
in the loop integrals. The massless limits are
actually
smooth, and arise often in practice. It is therefore useful to have
explicit expressions for
those massless limits that are not immediately obvious. Using
eqs.~(\ref{deltaforce})-(\ref{sealteamsix}), they
are found to be:
\beq
\fSSV(x,y,0) &=& (x+y)^2 + 3 (x+y) \ihat(x, y,0) +3 \jhat(x,y) 
- 2 x J(x) - 2 y J(y)
  ,\\
\fVVS (x,0,z) &=& 
-{3x\over 4}  - {z\over 2} - {3 z\over 4 x} \ihat (0, 0,z) 
+ ({3z \over 4 x} - {9 \over 4}) \ihat (0,x, z) 
+ {3 \over 4 x} \jhat (x, z) 
\nonumber \\ && 
+ 2 J(z) 
  ,\\
\fVVS (0,0,z) &=& 
- 3 \ihat (0,0,z) + {7\over 2} J(z)
-{5 z \over 4} 
  ,\\
\fFFV (x,y,0) &=& 0 
  ,\\
\fgauge (x,y,0) &=& {1\over 4 x y} \Bigl \lbrace
(43 x^2 y + 43 x y^2
-7 x^3 - 7 y^3 ) \ihat (0,x,y)
\nonumber \\ &&
+(2 y + 7 x) x^2 \ihat (0,0,x) 
+(2 x + 7 y) y^2 \ihat (0,0,y) 
\nonumber \\ &&
+(34 x y -7 x^2 -7 y^2) \jhat (x,y) 
\Bigr \rbrace 
+ 4 x^2 +4 y^2 +{35\over 2} x y 
\nonumber \\ &&
-{19\over 3} [x J(x) + y J(y)] 
+ 5 [y J(x) + x J(y)] 
  , \\
\fgauge (x,0,0) &=& 
13 x \ihat (0,0,x) 
-{59 \over 6} x J(x) 
+ {23\over 4} x^2 
.
\eeq
All of the functions vanish (by dimensional analysis) whenever all
arguments vanish.

It may also be of interest to see the individual contributions of the
Feynman diagrams labelled $VV$, $VVV$, and $ggV$ in figure
\ref{fig:feynmandiagrams}, even though these can always be combined
into $V^{(2)}_{\rm gauge}$. These
contributions are listed in Appendix A.

The classic results of Ford, Jack and Jones for the Standard Model
\cite{Ford:1992pn} are
a particularly useful special case of those found in this 
section, with which I have checked agreement. In fact, each type of term
that can occur in a
general
model in $\overline{\rm MS}$ does in fact arise in the Standard Model
case; no new types
of integrals arise,
so that the results of eqs.~(\ref{fSSS})-(\ref{fgaugeMSbar}) could be
inferred
from \cite{Ford:1992pn} by some forensic combinatorics.
Their
functions $A(x,y,z)$, $B(x,y,z)$,
$C(x,y)$, $D(x,y,z)$, $E(x,y)$, $\Sigma (x,y)$,
and $\Delta (x,y,z)$ are respectively
equal to the functions $\fSSV (x,y,z)$, $-\fVVS (x,y,z)$, $\fVS (x,y)$,
$-\fFFV (x,y,z)$, $\sqrt{xy}\fFFbV (x,y,z)$, $\fVV (x,y)$, and $-\fVVV
(x,y,z)$
given in this section and in Appendix A.
[Note that after the published errata of ref. \cite{Ford:1992pn}, a few
further minor
typographical errors have been recently corrected in the eprint archive
version.]

\section{Two-loop effective potential in the 
$\overline{\rm DR}$ scheme}
\label{sec:drbar}
\setcounter{equation}{0}
\setcounter{footnote}{1}

In this section, I report the results for the effective potential in the
$\overline{\rm DR}$ scheme. These are obtained by keeping all
4 components of each vector field, but performing momentum integrations
in $4 - 2 \epsilon$ dimensions. The difference, compared to the results
for $\overline{\rm MS}$, 
can be organized in terms of the extra epsilon scalars with multiplicity
$2 \epsilon$. Of course, the $SSS$, $SS$, $FFS$, and $\FFbS$ diagrams
in fig.~\ref{fig:feynmandiagrams} are unaffected by this procedure. Also,
the $SSV$ and $ggV$ diagrams are
unchanged in going from $\overline{\rm MS}$ to $\overline{\rm DR}$,
because in those cases all the vector indices are contracted with a
loop momentum.
The $VS$, $FFV$ and $\FFbV$ diagrams yield new 
contributions, which we can call $\epsilon S$,
$FF\epsilon$ and $\FFbe$, when the vector line in each case is turned into
an epsilon-scalar line.  
In the VVS diagram, a non-vanishing additional contribution arises
only when both vectors
are turned into epsilon scalars; call this contribution $\epsilon\epsilon
S$.
In the $VV$ diagram, one or both of the vector lines
can become an epsilon scalar, yielding contributions to be called
$\epsilon V$ and $\epsilon\epsilon$ respectively. Finally, in the $VVV$
diagram, any two of the vector lines can be turned into 
epsilon-scalar lines, resulting in a contribution $\epsilon\epsilon V$.

As discussed in section \ref{sec:oneloop}, the couplings of epsilon
scalars
have exactly the form indicated for vectors in
eqs.~(\ref{LSV})-(\ref{Lgauge}),
except that
when an epsilon scalar is involved, the 
rotation to the squared-mass eigenstate basis requires
$N^{(\epsilon )}$ rather than $N^{(V)}$.
This distinction is indicated by replacing the vector index 
$a,b,c,\ldots$
by an epsilon-scalar index $\hat a, \hat b, \hat c,\ldots$ on the
couplings.
For example [compare to eqs.~(\ref{definegabc})-(\ref{definegapq})],
the epsilon scalar-epsilon scalar-vector,
epsilon scalar-vector-vector,
and fermion-fermion-epsilon scalar couplings are
\beq
g^{\ahat b c} &=& g f^{efg} 
N^{(\epsilon)}_{ae}
N^{(V)}_{bf}
N^{(V)}_{cg} , 
\\
g^{\ahat \bhat c} &=& g f^{efg} 
N^{(\epsilon)}_{ae}
N^{(\epsilon)}_{bf}
N^{(V)}_{cg} ,
\\
g^{\hat a J}_I &=& g (T^b)^K_L N^{(F)*}_{JK} N^{(F)}_{IL}
N^{(\epsilon)}_{ab}
.
\eeq
Then the result in the $\overline{\rm DR}$ scheme can be written
\beq
V^{(2)}_{\overline{\rm DR}} &=& 
V^{(2)}_{\overline{\rm MS}}  
+ V^{(2)}_{\epsilon S}
+ V^{(2)}_{\epsilon\epsilon S}
+ V^{(2)}_{FF\epsilon}
+ V^{(2)}_{\FFbe}
+ V^{(2)}_{\epsilon V}
+ V^{(2)}_{\epsilon\epsilon}
+ V^{(2)}_{\epsilon\epsilon V}   ,
\eeq
where
\beq
V^{(2)}_{\epsilon S} &=& 
{1\over 4} g^{\ahat\ahat ii} \feS (\hat m^2_{a}, m^2_i) 
  ,\\
V^{(2)}_{\epsilon\epsilon S} &=& {1\over 4} (g^{\ahat\bhat i})^2 \feeS
(\hat m^2_{a}, \hat m^2_{b}, m^2_i)
  ,\\
V^{(2)}_{FF\epsilon} &=& {1\over 2} |g^{\ahat J}_I|^2 \fFFe
(m^2_I, m^2_J, \hat m^2_{a})
  ,\\
V^{(2)}_{\FFbe} &=& {1\over 2} g^{\ahat J}_I g^{\ahat J'}_{I'} 
M^{II'} M^*_{JJ'} \fFFbe (m^2_I, m^2_J,\hat m^2_{a} )
  ,\\
V^{(2)}_{\epsilon V} &=& {1\over 2} (g^{\ahat b c})^2 \feV 
(\hat m^2_{a},m^2_b)
  ,\\
V^{(2)}_{\epsilon \epsilon} &=& {1\over 4} (g^{\ahat \bhat c})^2 \fee 
(\hat m^2_{a},\hat m^2_{b})
  ,\\
V^{(2)}_{\epsilon\epsilon V} &=& {1\over 4} (g^{\ahat\bhat c})^2 \feeV 
(\hat m^2_{a},\hat m^2_{b},m^2_c)
  ,
\eeq
with the loop functions given by:
\beq
\feS (x,y) &=& - 2 x J(y)
  ,\\
\feeS (x,y,z) &=& -2 J(z) +x+y+z
  ,\\
\fFFe (x,y,z) &=& 2 x J(x) +2 y J(y) -(x+y)^2 +z^2  
  ,\\
\fFFbe (x,y,z) &=& 4 J(x) + 4 J(y) - 2 x - 2 y - 2 z
  ,\\
\feV (x,y) &=& -4 x y - 6 x J(y)
  ,\\
\fee (x,y) &=& 4 x y
  ,\\
\feeV (x,y,z) &=& -x^2 - y^2 +z^2 -6 x y  -x z -y z + (6x+6y-2z)J(z) 
  .
\eeq
This completes the result for the two-loop effective potential in the
$\overline{\rm DR}$ scheme.
\section{Two-loop effective potential in the 
$\overline{\rm DR}'$ scheme}
\label{sec:drbarprime}
\setcounter{equation}{0}
\setcounter{footnote}{1}

As explained in the Introduction and in section \ref{sec:oneloop},
it is convenient in models of softly broken supersymmetry 
to go to the $\overline{\rm DR}'$
scheme. This scheme is defined so that $m_{\epsilon}^2$ (the difference
between the squared
masses of epsilon
scalars and their vector counterparts) does not appear in the
beta functions of other couplings, or in the effective potential,
or in the equations relating pole masses to running masses.
Starting from the $\overline{\rm DR}$ results of the previous section,
I find that this is done at two-loop order 
by the following parameter redefinition of
soft terms appearing in eq.~(\ref{softterms}):
\beq
(m^2_{\overline{\rm DR}'})_i^j &=& (m^2_{\overline{\rm DR}})_i^j 
-{1\over 16 \pi^2} \left [ \delta_i^j 2 g^2 C(i)  
m_{\epsilon}^2
\right ] 
+ {1\over (16 \pi^2)^2}\Bigl \lbrace 
Y^{ikl} Y_{jkl} g^2 [C(k) - {1\over 2} C(i)] 
m_\epsilon^2
\nonumber \\ && \qquad\qquad\qquad\qquad
+\delta_i^j g^4 C(i) \left [ 2 S(R) + 4 C(i) - 6 \CG
\right ]
m_\epsilon^2
\Bigr \rbrace
  ,
\\
c^i_{\overline{\rm DR}'} &=& 
c^i_{\overline{\rm DR}} + {1\over (16 \pi^2)^2}
\left [ Y^{ijk} \mu_{jk} g^2 C(j) m_\epsilon^2 \right ]
  ,
\\
\Lambda_{\overline{\rm DR}'} &=& \Lambda_{\overline{\rm DR}}
- {1\over 16 \pi^2} {d_G (m_\epsilon^2)^2\over 2}
+ {1\over (16 \pi^2)^2}\Bigl \lbrace 
{g^2\over 2} d_G [S(R)- \CG] (m_\epsilon^2)^2
\nonumber \\  && \qquad\qquad\qquad\qquad\qquad
+ g^2 d_G \CG |M|^2 m_\epsilon^2 + g^2 \mu^{ij} \mu_{ij} C(i)
m_\epsilon^2
\Bigr \rbrace
  .
\eeq
If there is more than one simple or $U(1)$ group, then each of the
correction terms should be summed over subgroups, 
with a different $m_\epsilon^2$ for each subgroup.
The exception is that the term
\beq
g^4 C(i)^2 \rightarrow \sum_a \sum_b g_a^2 g_b^2 C_a(i) C_b(i),
\label{Osamamustdie}
\eeq
involves a double sum over subgroups labeled $a,b$.
Following these redefinitions, the result
for the full two-loop effective potential turns out to have the same
functional
form as if one
naively took the $\overline{\rm DR}$ result and set $m^2_\epsilon$ to 0,
removing the distinction between $N^{(\epsilon )}$ and $N^{(V)}$, 
between hatted and unhatted vector squared-mass
eigenstate indices on the couplings, and between $\hat m^2_a$ and
$m^2_a$.
It is therefore convenient to define functions which combine the
effects of the $4-2 \epsilon$ vectors and the epsilon scalars:
\beq
\FVS (x,y) &=& \fVS (x,y) + \feS (x,y) 
  ,\\
\FVVS (x,y,z) &=& \fVVS (x,y,z) + \feeS (x,y,z) 
  ,\\
\FFFV (x,y,z)  &=& \fFFV (x,y,z) + \fFFe (x,y,z) 
  ,\\
\FFFbV (x,y,z) &=& \fFFbV (x,y,z) + \fFFbe (x,y,z)
  ,\\
\Fgauge (x,y,z) &=& \fgauge (x,y,z) + \feeV (x,y,z) + \feeV (z,x,y) +
\feeV (y,z,x) 
\nonumber \\ &&
+ \feV (x,y) 
+ \feV (y,x) 
+ \feV (x,z) 
+ \feV (z,x) 
+ \feV (y,z) 
+ \feV (z,y) 
\nonumber \\ &&
+ \fee (x,y) + \fee (x,z) + \fee (y,z) 
  .
\eeq
Note that I use $F$'s rather than $f$'s to distinguish the $\overline{\rm
DR}'$ functions from the corresponding $\overline{\rm MS}$ functions.

Therefore, the 
$\overline{\rm DR}'$ 
two-loop effective potential is given
by:
\beq
V^{(2)} &=&
V^{(2)}_{SSS}
+ V^{(2)}_{SS}
+ V^{(2)}_{FFS}
+ V^{(2)}_{\FFbS}
+ V^{(2)}_{SSV}
+ V^{(2)}_{VS}
+ V^{(2)}_{VVS}
\nonumber \\ &&
+ V^{(2)}_{FFV}
+ V^{(2)}_{\FFbV}
+ V^{(2)}_{\rm gauge}
  ,
\eeq
where now
\beq
V^{(2)}_{SSS} &=& {1\over 12} (\lambda^{ijk})^2
                  \fSSS (m^2_i, m^2_j, m^2_k)
  ,
\label{VSSSDrbarprime}
\\
V^{(2)}_{SS} &=& {1\over 8} \lambda^{iijj} \fSS (m^2_i, m^2_j) 
  ,\\
V^{(2)}_{FFS} &=& {1\over 2} |y^{IJk}|^2 \fFFS (m^2_I, m^2_J, m^2_k)
  ,\\  
V^{(2)}_{\FFbS} &=& {1\over 4} y^{IJk} y^{I'J'k} M^*_{II'} M^*_{JJ'} 
                             \fFFbS (m^2_I, m^2_J, m^2_k) + {\rm c.c.}
  ,\\
V^{(2)}_{SSV} &=& {1\over 4} (g^{aij})^2 \fSSV (m^2_i,m^2_j, m_a^2)
  ,\\
V^{(2)}_{VS} &=& {1\over 4} g^{aaii} \FVS (m^2_a, m^2_i) 
  ,\\ 
V^{(2)}_{VVS} &=& {1\over 4} (g^{abi})^2 \FVVS (m^2_a, m^2_b,m^2_i) 
  ,\\
V^{(2)}_{FFV} &=& {1\over 2} |g^{aJ}_I|^2 \FFFV (m^2_I, m^2_J, m^2_a)
  ,\\
V^{(2)}_{\FFbV} &=& {1\over 2} g^{aJ}_I g^{aJ'}_{I'}
M^{II'} M^*_{JJ'} \FFFbV (m^2_I, m^2_J, m^2_a)
  ,\\
V^{(2)}_{\rm gauge} &=& {1\over 12} (g^{abc})^2 \Fgauge
(m^2_a,m^2_b,m^2_c)
  .
\label{VgaugeDRbarprime}
\eeq
Here $\fSSS (x,y,z)$, $\fSS (x,y)$, $\fFFS(x,y,z)$, 
$\fFFbS (x,y,z)$, and $\fSSV (x,y,z)$ are given by exactly the same
functions as in $\overline{\rm MS}$, eqs.~(\ref{fSSS})-(\ref{fSSV}).
The new functions are given by:
\beq
\FVS (x,y) &=& 3 \jhat (x,y)   ,
\label{FVS}
\\
\FVVS (x,y,z) &=& {1 \over 4 x y}  \Bigl \lbrace
       (-x^2 - y^2 -z^2 -10 x y +2 x z +2 y z) \ihat (x,y,z) 
\nonumber \\ && 
       +(x-z)^2 \ihat (0,x,z) 
+(y-z)^2 \ihat (0,y,z) 
       - z^2 \ihat (0,0,z) 
\nonumber \\ && 
+(z-x-y) \jhat (x,y) +y \jhat(x,z) +x \jhat (y,z) 
\Bigr \rbrace
\nonumber 
\\ 
&& 
+ {1\over 2}J(x) + {1\over 2} J(y) ,
\\
\FFFV (x,y,z) &=& 
   {1\over z} \Bigl \lbrace (x^2 +y^2 -2 z^2 -2 x y+x z+y z) 
   \ihat (x,y,z) 
   - (x-y)^2 \ihat (0,x,y) 
   \nonumber \\ && 
   +(x-y-2z) \jhat (x,z) + (y-x-2 z) \jhat (y,z) 
   +2 z \jhat (x,y) 
   \Bigr \rbrace 
   \nonumber \\ &&                 
   + 2 (-x - y + z/3) J(z)
  ,\\
\FFFbV (x,y,z) &=& 6 \ihat(x,y,z)
  ,\\
\Fgauge (x,y,z) &=& {1\over 4 x y z} \Bigl \lbrace
(-x^4 - 8 x^3 y - 8 x^3 z 
+32 x^2 y z +18 y^2 z^2 
) \ihat (x,y,z)
\nonumber \\ &&
+ (y-z)^2 (y^2 + 10 y z + z^2) \ihat (0,y,z) 
+ x^2 (2 y z - x^2) \ihat (0,0,x)
\nonumber \\ && 
+ (x^2 - 9 y^2 - 9 z^2 + 9 x y + 9 x z +14 y z) x \jhat(y,z)
\nonumber \\ && 
+ (22 y + 22 z- 40x/3) xyz J(x)
\Bigr \rbrace 
\nonumber \\ &&
+ (x \leftrightarrow y) + (x \leftrightarrow z)
  .
\label{Fgauge}
\eeq

Despite the appearance of $x,y,z$ in the
denominators, these functions again all 
have smooth limits for $x,y,z \rightarrow 0$. The non-trivial ones are
\beq
\FVVS (x,0,z) &=& 
{x\over 4}  + {z\over 2} - {3 z\over 4 x} \ihat (0, 0,z) 
+ ({3z \over 4 x}-{9 \over 4} ) \ihat (0,x, z) 
+ {3 \over 4 x} \jhat (x, z) 
  ,
\label{FVVSx0z}
\\
\FVVS (0,0,z) &=& - 3 \ihat (0,0,z) + {3\over 2} J(z)-{z \over 4}
  ,\\
\FFFV (x,y,0) &=& -(x+y)^2 + 2 x J(x) + 2 y J(y) 
  ,\\
\Fgauge (x,y,0) &=& {1\over 4 x y} \Bigl \lbrace
(43 x^2 y + 43 x y^2 -7 x^3 - 7 y^3 ) \ihat (0,x,y)
\nonumber \\ &&
+(2 y + 7 x) x^2 \ihat (0,0,x) 
+(2 x + 7 y) y^2 \ihat (0,0,y) 
\nonumber \\ &&
+(34 x y - 7 x^2 -7 y^2 ) \jhat (x,y) 
\Bigr \rbrace 
+ 3 x^2 +3 y^2 +{11\over 2} x y 
\nonumber \\ &&
-{25\over 3} [x J(x) + y J(y)] 
+ 5 [y J(x) + x J(y)] 
  ,\\
\Fgauge (0,0,x) &=& 
13 x \ihat (0,0,x) 
- {71 \over 6} x J(x)
+{19\over 4} x^2 
  .
\label{Fgauge00x}
\eeq
Also, it may be of interest to see the contributions from individual
graphs to $\Fgauge (x,y,z)$. Those are listed in Appendix A.

This completes the result for the two-loop effective potential in
the $\overline{\rm DR}'$ scheme. These are appropriate for use in
any softly broken supersymmetric model, including the MSSM. Partial
results for the MSSM corresponding to the leading contributions 
proportional to
$\alpha_S y_t^2$, $\alpha_S y_b^2$, and 
$y^4_t$, $y_b^4$ have been given in refs.~\cite{Zhang:1999bm}
and
\cite{Espinosa:2000df}.
Several illustrative examples and
consistency checks are done  in section \ref{sec:examples}.

\section{Renormalization group invariance of the two-loop effective
potential in softly broken supersymmetry}
\label{sec:rginv}
\setcounter{equation}{0}
\setcounter{footnote}{1}

In general, the condition for RG invariance of the
effective
potential is
\beq
Q {d V \over dQ} = \left ( Q {\partial \over \partial Q} 
+ \sum_I \beta_{\lambda_I} {\partial \over \partial \lambda_I} - 
\sum_i \gamma_i^{(S)} \phi_i {\partial \over \partial \phi_i}
\right ) V = 0 .
\label{rgcheck}
\eeq
Here, $\lambda_I$ are all of the running parameters of the 
model with beta functions $\beta_{\lambda_I}$, and $\gamma^{(S)}_i$ are
the
anomalous dimensions of the scalar fields $\phi_i$.
At one- and two-loop order, this means
\beq
&& Q{\partial \over \partial Q} V^{(1)} 
 + 
\left [ 
\sum_I \beta_{\lambda_I}^{(1)} {\partial \over \partial \lambda_I} 
-\sum_i \gamma_i^{(S,1)} \phi_i {\partial \over \partial \phi_i} 
\right ] 
V^{(0)}
\> = \> 0
  ,
\label{oneloopRGcheck}
\\
&& Q{\partial \over \partial Q} V^{(2)} 
 +  \left[ 
\sum_I \beta_{\lambda_I}^{(1)} {\partial \over \partial \lambda_I} 
-\sum_i \gamma_i^{(S,1)} \phi_i {\partial \over \partial \phi_i} 
\right ] V^{(1)} 
 \nonumber \\ 
&&\qquad\qquad \>\>+  
\left [ 
\sum_I \beta^{(2)}_{\lambda_I} {\partial \over \partial \lambda_I} 
-\sum_i \gamma_i^{(S,2)} \phi_i {\partial \over \partial \phi_i}
\right ] 
V^{(0)} 
\> = \> 0
  .
\label{twoloopRGcheck}
\eeq

In softly broken supersymmetry, I find that the anomalous dimension matrix
for
scalar fields in the Landau gauge and in either
$\overline{\rm DR}$ or $\overline{\rm DR}'$ is
\beq
\gamma^{(S)j}_i &=& 
{1\over 16 \pi^2} \gamma_i^{(S,1)j}
+ {1\over (16 \pi^2)^2} \gamma_i^{(S,2)j} 
  ;
\\
\gamma_i^{(S,1)j} &=& 
{1\over 2} Y_{ikl} Y^{jkl} -  \delta_i^j g^2 C(i) 
, 
\label{gammaSone}
\\
\gamma_i^{(S,2)j} &=& 
- {1\over 2} Y_{imn} Y^{nkl} Y_{klr} Y^{mrj} 
+ Y_{ikl} Y^{jkl} g^2 [ 2 C(k) - C(i)]
\nonumber \\ &&
+ \delta_i^j g^4 C(i) [S(R) + 2 C(i) - {9\over 4} \CG ] .
\label{gammaStwo}
\eeq
This can be obtained starting from the general results in the 
$\overline{\rm MS}$ scheme in
ref.~\cite{Machacek:1983tz}, and then applying the coupling constant
redefinitions needed to transform from the 
$\overline{\rm MS}$ to the $\overline{\rm DR}$ or $\overline{\rm DR}'$
scheme \cite{Martin:1993yx}. The eigenvalues of this matrix
constrained to the subspace of the classical scalar background fields give
the anomalous dimensions appearing in eqs.~(\ref{oneloopRGcheck})
and (\ref{twoloopRGcheck}).
It should be noted that because of gauge-fixing, the Landau gauge scalar
field anomalous dimension matrix  
$\gamma^{(S)j}_i$ relevant for the effective potential is
{\it not} the
same
as the more widely-known, gauge-invariant, anomalous dimension matrix of 
the 
chiral 
superfields. For
comparison, the latter is \cite{anomalousdimension}
\beq
\gamma^{j}_i &=& 
{1\over 16 \pi^2} \gamma_i^{(1)j}
+ {1\over (16 \pi^2)^2} \gamma_i^{(2)j} ;
\\
\gamma_i^{(1)j} &=& 
{1\over 2} Y_{ikl} Y^{jkl} - 2 \delta_i^j g^2 C(i) 
, 
\label{gammaone}
\\
\gamma_i^{(2)j} &=& 
- {1\over 2} Y_{imn} Y^{nkl} Y_{klr} Y^{mrj} 
+ Y_{ikl} Y^{jkl} g^2 [ 2 C(k) - C(i)]
\nonumber \\ &&
+ \delta_i^j g^4 C(i) [2 S(R) + 4 C(i) - 6 \CG ] .
\label{gammatwo}
\eeq

In order for the effective potential to satisfy eq.~(\ref{rgcheck})
in a model with explicit supersymmetry breaking, it is necessary to
include a running vacuum energy term $\Lambda$, as in
eq.~(\ref{softterms}).
Now using the results of section \ref{sec:drbarprime}, one can derive
the $\overline{\rm DR}'$ beta function for $\Lambda$ up to two loops in a
general
softly-broken supersymmetric model as specified in subsection
\ref{subsec:soft}, by looking at the $\phi_i$-independent piece
of eqs.~(\ref{oneloopRGcheck})-(\ref{twoloopRGcheck}). I find
\beq
\beta_\Lambda &=& {1\over 16 \pi^2} \beta^{(1)}_\Lambda + 
{1\over (16 \pi^2)^2} \beta^{(2)}_\Lambda \\
\beta^{(1)}_\Lambda &=& 
(m^2)_i^j (m^2)_j^i +
2 (m^2)_i^j \mu^{ik} \mu_{kj} + b^{ij} b_{ij} - d_G |M|^4
, 
\label{betaLambdaone}
\\
\beta^{(2)}_\Lambda &=&
g^2 d_G |M|^4 [ 4 S(R) - 8 \CG ]  
+8 g^2 |M|^2 \mu^{ij} \mu_{ij} C(i)
+ 8 g^2 (m^2)_i^j \mu^{ik} \mu_{kj} C(i)
\nonumber \\  &&
+ 4 g^2 (m^2)_i^j (m^2)_j^i C(i)
+ 4 g^2 b^{ij} b_{ij} C(i)
- 4 g^2 M \mu^{ij} b_{ij} C(i) 
- 4 g^2 M^* \mu_{ij} b^{ij} C(i) 
\nonumber \\  &&
-Y^{ijk} Y_{ijl} \left [ 
(m^2)_k^m (m^2)_m^l 
+ (m^2)_k^m \mu_{mn} \mu^{nl}
+ \mu_{km} \mu^{mn} (m^2)_n^l
+ \mu_{km} (m^2)_n^m \mu^{nl}
+ b_{km} b^{ml} \right ]
\nonumber \\  &&
- a^{ijk} a_{ijl} \left [ (m^2)_k^l + \mu_{km} \mu^{ml} \right ]
-2 Y^{ijk} Y_{ilm} (m^2)_j^l \mu^{mn} \mu_{nk}
\nonumber \\  &&
- Y^{ijk} a_{ijl} \mu_{km} b^{ml}
- Y_{ijk} a^{ijl} \mu^{km} b_{ml} .
\label{betaLambdatwo}
\eeq
where $d_G$ is the dimension of the adjoint representation of the group.
If the gauge group contains more than one simple or $U(1)$ component,
then terms involving the gaugino mass $M$ or $g^2$ should be summed over
subgroups in 
eqs.~(\ref{gammaSone})-(\ref{gammaStwo}),
(\ref{gammaone})-(\ref{gammatwo}),
and (\ref{betaLambdaone})-(\ref{betaLambdatwo}),
with the exception noted in eq.~(\ref{Osamamustdie}).
Special cases of these general results will be used in the next section.

I have checked explicitly that the $\overline{\rm DR}'$ two-loop effective
potential for a general softly-broken supersymmetric theory satisfies RG
invariance, using the results given above and in Appendix B and in
refs.~\cite{Martin:1994zk,Jack:1994rk}. 

\section{Examples and consistency checks}
\label{sec:examples}
\setcounter{equation}{0}
\setcounter{footnote}{1}

In this section, I study some examples chosen as consistency checks and
useful points of reference for the results given above. The examples are
all based on supersymmetry with or without soft breaking, so the
$\overline{\rm DR}'$ scheme is used. One type of consistency check follows
from the requirement that the two-loop effective potential satisfies
RG invariance in conjunction with the known two-loop
beta functions \cite{Martin:1994zk,Yamada:1994id,Jack:1994kd,Jack:1994rk},
and the scalar
anomalous dimensions and $\beta_\Lambda$ found in the previous section. 
(Since the result for $\beta_\Lambda$ is itself a consequence of the 
calculation, that part is not really an independent check.)
The derivatives of the loop functions
are listed in Appendix B.
Another type of
check relies on the fact that the effective potential for a supersymmetric
theory in a supersymmetric vacuum must vanish. These
consistency checks rely on non-trivial cancellations between different
two-loop functions, which are made manifest by writing them in terms of
the basis functions $\ihat(x,y,z)$, $\jhat (x,y)$, and $J(x)$,
using eqs.~(\ref{fSSS})-(\ref{fSSV}) and (\ref{FVS})-(\ref{Fgauge00x}).

\subsection{The Wess-Zumino Model}
\label{subsec:wz}

Consider the Wess-Zumino model \cite{WessZumino} with a single chiral
supermultiplet $\Phi$ containing a Weyl fermion $\psi$ and a complex
scalar $
\phi + {(R + i I)/\sqrt{2}}, 
$
where $\phi$ is the classical background, and $R$, $I$ are real
scalar quantum
fluctuations. The
superpotential is given by
\beq
W = {\mu\over 2} \Phi^2 + {y\over 6 } \Phi^3 
,
\eeq
where $\mu$ and $y$ are mass and coupling parameters, taken to be
real without loss of generality. 
The fields $R$, $I$, $\psi$ are mass eigenstates, with
\beq
m_R^2 &=& \mu^2 + 3  y\mu \phi + 3y^2 \phi^2/2
,\\
m_I^2 &=& \mu^2 + y \mu \phi + y^2 \phi^2/2,\\
m_\psi &=& \mu + y \phi .
\eeq
The tree-level scalar potential is
\beq
V^{(0)} = \mu^2 \phi^2 + y \mu \phi^3 + y^2 \phi^4/4
,
\label{VtreeWZ}
\eeq
and the one-loop contribution to the effective potential is given in
terms of the function $\oneloop (x)$ in eq.~(\ref{defineoneloop}) by
\beq
V^{(1)} = \oneloop (m_R^2) + \oneloop (m_I^2) - 2 \oneloop (m_\psi^2)
.
\label{oneloopVWZ}
\eeq
The non-zero scalar quartic and cubic couplings are:
\beq
\lambda^{RRRR} = \lambda^{IIII} &=& 3y^2/2,\\
\lambda^{RRII} &=& y^2/2 ,\\
\lambda^{RRR} &=& 3 y (\mu+ y \phi)/\sqrt{2},\\
\lambda^{RII} &=& y (\mu+ y \phi)/\sqrt{2},
\eeq
and the Yukawa interactions are
\beq
y^{\psi\psi R} &=& y/\sqrt{2},\\
y^{\psi\psi I} &=& iy/\sqrt{2} 
.
\eeq
It follows that the contributions to the two-loop effective potential are:
\beq
V_{SSS}^{(2)} &=& 
{y^2 \over 8} (\mu + y \phi)^2 \left [3 \fSSS (m^2_R, m^2_R, m^2_R) 
+ \fSSS (m^2_R, m^2_I, m^2_I) \right ]
,
\label{wzsss}
\\
V_{SS}^{(2)} &=& 
{y^2 \over 16} \left [ 3 \fSS (m^2_R, m^2_R) +
3 \fSS (m^2_I, m^2_I) +
2 \fSS (m^2_R, m^2_I) \right ]
,
\label{wzss} 
\\
V_{FFS}^{(2)} &=&
{y^2 \over 4} \left [ \fFFS (m_\psi^2, m_\psi^2, m^2_R) +
\fFFS (m_\psi^2, m_\psi^2, m^2_I) \right ]
,
\label{wzffs}
\\
V_{\FFbS}^{(2)} &=& {y^2 \over 4} m_\psi^2 
\left [ \fFFbS (m_\psi^2, m_\psi^2, m_R^2) - \fFFbS (m_\psi^2, m_\psi^2,
m_I^2)\right ].
\label{wzffbs}
\eeq

Now one may check RG invariance of the effective
potential. At one-loop order, one finds from eq.~(\ref{oneloopVWZ}) that
\beq
Q{\partial \over \partial Q} V^{(1)} = 
-y^2 \mu^2 \phi^2  
-y^3 \mu \phi^3  
-y^4 \phi^4/4 .
\eeq
The one-loop scalar anomalous dimension and beta functions are
\beq
\gamma_\phi^{(S,1)} &=& y^2/2 ,\\
\beta_\mu^{(1)} &=& y^2 \mu ,\\
\beta_y^{(1)} &=& 3y^3/2 .
\eeq
Therefore, from eq.~(\ref{VtreeWZ}):
\beq
\sum_I \beta_{\lambda_I}^{(1)} {\partial \over \partial \lambda_I} 
V^{(0)} &=& 
2 y^2 \mu^2 \phi^2 
+ 5 y^3 \mu \phi^3/2
+ 3 y^4 \phi^4/4 
, \\
-\gamma_\phi^{(S,1)} \phi {\partial \over \partial \phi} V^{(0)} &=& 
-y^2 \mu^2 \phi^2- 3 y^3 \mu \phi^3/2
-y^4 \phi^4/2,
\eeq
where $\lambda_I$ runs over $y,\mu$, so that
eq.~(\ref{oneloopRGcheck}) is indeed satisfied.
At two-loop order, one finds from 
eqs.~(\ref{wzsss})-(\ref{wzffbs}) and (\ref{dQfsss})-(\ref{dQfffbs})
that
\beq
Q{\partial \over \partial Q} V^{(2)} 
+ \left [ \sum_I\beta_{\lambda_I}^{(1)} {\partial \over \partial
\lambda_I} 
- \gamma_\phi^{(S,1)} \phi {\partial \over \partial \phi} \right ]
V^{(1)} =
y^4 \mu^2 \phi^2
+ y^5 \mu \phi^3 
+y^6 \phi^4/4 
.
\label{WZcheckb1}
\eeq
From the two-loop RG scalar anomalous dimension and beta functions:
\beq
\gamma_\phi^{(S,2)} &=& -y^4/2
,\\
\beta_\mu^{(2)} &=& -y^4 \mu
,\\
\beta_y^{(2)} &=& -3y^5/2 
,
\eeq
one also finds:
\beq
\sum_I \beta_{\lambda_I}^{(2)} {\partial \over \partial \lambda_I}
V^{(0)} &=&
- 2 y^4 \mu^2 \phi^2
- 5 y^5 \mu \phi^3/2
- 3 y^6 \phi^4/4 
,
\label{WZcheckb2}
\\
-\gamma_\phi^{(S,2)} \phi {\partial \over \partial \phi} V^{(0)} &=& 
y^4 \mu^2 \phi^2 + 3 y^5 \mu \phi^3/2 + y^6  \phi^4/2
.
\label{WZcheckb3}
\eeq
The results of eqs.~(\ref{WZcheckb1}),
(\ref{WZcheckb2}) and
(\ref{WZcheckb3})
combine to verify eq.~(\ref{twoloopRGcheck}).

In the special case of $\phi=0$, supersymmetry is not broken,
and the effective potential should vanish. At
one-loop order, eq.~(\ref{oneloopVWZ})
then vanishes trivially. At two-loop order,
\beq
V^{(2)} = {y^2 \over 2} \left [
\mu^2 \fSSS (\mu^2,\mu^2,\mu^2) 
+ \fSS (\mu^2, \mu^2) 
+ \fFFS (\mu^2, \mu^2, \mu^2)
\right ] ,
\eeq
which equals 0 by virtue of eqs.~(\ref{fSSS})-(\ref{fFFS}).

\subsection{Supersymmetric QED in supersymmetric vacua}
\label{subsec:SQED}

Let us now consider a supersymmetric $U(1)$ gauge theory with
coupling constant $g$ and a pair of chiral superfields with charges $\pm
1$.

First take the case that the chiral superfields do not have a mass
term before symmetry breaking, and the two scalar fields have the same
classical background value $\phi$.
Then the gauge symmetry is broken, but supersymmetry remains
unbroken since $\phi$ parameterizes a flat direction. 
The vector
boson, two Weyl fermions, and a real scalar field each obtain a mass 
\beq
x = 4 g^2 \phi^2. 
\eeq 
Together with a massless (in Landau gauge) real
scalar Nambu-Goldstone boson, these form a massive vector supermultiplet.
In
addition, there are two massless real scalars and one massless Weyl
fermion forming a chiral supermultiplet. 
The $\overline{\rm DR}'$ one-loop effective potential
vanishes because of these mass degeneracies.
The two-loop effective potential contributions
in the $\overline{\rm DR}'$ scheme are:
\beq
V^{(2)}_{SSS} &=& g^2 x \left [ {1\over 2} \fSSS (0,0,x) + \fSSS
(0,x,x)
\right ] 
,
\label{sqedsss} \\
V^{(2)}_{FFS} &=& g^2 \left [ \fFFS (0,x,0) + \fFFS (0,x,x) + 2 \fFFS
(x,x,0) \right ] 
, \\
V^{(2)}_{SSV} &=& {g^2 \over 2} \left [ \fSSV (0,0,x) + \fSSV (0,x,x) 
\right ] 
, \\
V^{(2)}_{VS} &=& {g^2 \over 2} \FVS (x,x) 
,\\
V^{(2)}_{VVS} &=& g^2 x \FVVS (x,x,0) 
,\\
V^{(2)}_{FFV} &=& g^2 \FFFV (0,x,x) , \label{sqedffv}
\eeq
with the other contributions 
vanishing. One can now check
by plugging in the results of section \ref{sec:drbarprime} that the sum of
eqs.~(\ref{sqedsss})-(\ref{sqedffv}) yields 0, as required for a
supersymmetric vacuum. This constitutes a non-trivial identity
involving cancellations between different two-loop functions
which become apparent after writing them in terms of the
functions $\ihat (x,y,z)$, $\jhat (x,y)$ and $J(x)$.

Another check which relies on a different set of cancellations is
obtained if we take $\phi=0$ in the above model, but now include a
superpotential
mass term $\mu$. In that case, the vector gauge boson and the gaugino
are massless, and the real scalar fields and the chiral fermions all have
squared mass $\mu^2$. Then one obtains for the contributions to the
two-loop
effective potential in the $\overline{\rm DR}'$ scheme:
\beq
V^{(2)}_{SS} &=& g^2 \fSS (\mu^2, \mu^2) \label{sqedss} 
,\\
V^{(2)}_{FFS} &=& 4 g^2 \fFFS (0,\mu^2,\mu^2)  
,\\
V^{(2)}_{SSV} &=& g^2 \fSSV (\mu^2,\mu^2,0)
,\\
V^{(2)}_{FFV} &=& g^2 \FFFV (\mu^2,\mu^2,0) 
,\\
V^{(2)}_{\FFbV} &=& -g^2 \mu^2 \FFFbV (\mu^2,\mu^2,0) 
,
\label{sqedffbv} 
\eeq
with all other contributions vanishing. Again one finds from the
results of section \ref{sec:drbarprime} that the
sum of eqs.~(\ref{sqedss})-(\ref{sqedffbv}) yields 0, as required
for a supersymmetric vacuum.

\subsection{Supersymmetric $SU(\Nc)$ gauge theory with one
flavor in supersymmetric vacua}
\label{subsec:SuNc}

A richer set of checks is found in non-abelian supersymmetric
models. As an example, consider supersymmetric $SU(\Nc)$ gauge theory with
one flavor of chiral superfields $Q_i$ and $\overline Q^i$ in the
fundamental and anti-fundamental representations, respectively. Here
$i=1,\ldots ,\Nc$ is a color index. Consider evaluation of the effective
potential for the classical background:
\beq
\langle Q_i \rangle = \langle \overline Q^i \rangle 
= \delta_{i1} \phi
. 
\eeq
These VEVs break the gauge symmetry according to
$SU(\Nc) \rightarrow SU(\Nc-1)$, but $\phi$ parameterizes 
a flat direction
and supersymmetry is unbroken. Therefore the
effective potential must vanish at each order in perturbation theory
for any value of $\phi$. My aim is to show this explicitly.

The
particle content for non-zero $\phi$ consists of $2\Nc -1$ massive
vector supermultiplets with their associated massless (in Landau
gauge) real scalar Nambu-Goldstone modes,
$\Nc^2 - 2\Nc$ massless vector multiplets associated with the unbroken
gauge symmetry, and one massless singlet chiral supermultiplet.
The non-zero squared-mass eigenvalues are
\beq
x &=& g^2 \phi^2
,\\
y &=& {2(\Nc-1) \over \Nc} g^2 \phi^2,
\eeq
and the multiplicities of the mass eigenstates are shown in Table
\ref{multiplicities}.
\renewcommand{\arraystretch}{1.35}
\begin{table}[thb]
\caption{Multiplicities of mass eigenstates in the model of section
\ref{subsec:SuNc}.}
\label{multiplicities}
\begin{center}
\begin{tabular}
{|c||c|c|c|}
\hline
particle type & $m^2 = 0$ & $m^2 = x$ & $m^2 = y$ 
\\
\hline\hline
real scalars & $2\Nc+1$ & $2 \Nc-2$ & 1
\\
\hline
Weyl fermions & $\Nc^2-2\Nc+1$ & $4 \Nc-4$ & 2
\\
\hline
vectors & $\Nc^2 - 2 \Nc$ & $2 \Nc -2$ & 1
\\
\hline
\end{tabular}
\end{center}
\end{table}
Because of the mass degeneracies indicated in Table \ref{multiplicities},
the one-loop contribution to the effective potential vanishes as required.

At two-loop order, I find the contributions in the
$\overline{\rm DR}'$ scheme to be
\beq
V^{(2)}_{SSS} &=& g^4 \phi^2 \Biggl [
{\Nc -1 \over 4} \fSSS (0,0,x)
+ {(\Nc-1)^2 \over 2 \Nc^2} \fSSS (0,0,y)
+{\Nc - 1 \over 2} \fSSS (0,x,x)
\nonumber \\ &&
+ {(\Nc-1)^2 \over \Nc^2} \fSSS (0,y,y)
+ {(\Nc-2)^2 (\Nc-1)\over 4\Nc^2} \fSSS (0,x,y) 
\Biggr ]
,\\
V^{(2)}_{SS} &=& 0
,\\
V^{(2)}_{FFS} &=& 
g^2 \Biggl [ 
{2 \Nc^2 - 3 \Nc -1 \over 2} 
\left \lbrace \fFFS (0,x,0) + \fFFS (0,x,x) \right \rbrace
+ {3 (\Nc -1) \over 2} \fFFS (x,x,0) 
\nonumber \\ &&
+ {\Nc -1 \over 2 \Nc}
\left \lbrace \fFFS (0,y,0) + \fFFS (0,y,y) \right \rbrace
+ {\Nc-1 \over \Nc} \fFFS (y,y,0)
\nonumber \\ &&
+ {\Nc^2 - \Nc +2 \over 2 \Nc}
\left \lbrace \fFFS (x,y,0) + \fFFS (x,y,x) \right \rbrace
+ {\Nc -1 \over 2} \fFFS (x,x,y)
\Biggr ]
,\\
V^{(2)}_{\FFbS} &=& 
g^4 \phi^2 \Biggl [ 
{\Nc -1 \over 2}
\left \lbrace \fFFbS (x,x,y) - \fFFbS (x,x,0) \right \rbrace
\nonumber \\ &&
+ {2(\Nc -1)\over \Nc} \left \lbrace \fFFbS (x,y,x) 
- \fFFbS (x,y,0) 
\right \rbrace
\Biggr ] 
,\\
V^{(2)}_{SSV} &=& g^2 \Biggl [ 
{\Nc-1 \over 2} \fSSV (0,0,x) 
+ {1\over 4} \fSSV (0,0,y)
+ {\Nc -1 \over 4 \Nc} \fSSV (0,y,y)
\nonumber \\ && 
+ {\Nc -1 \over 4} \fSSV (0,x,x)
+ {\Nc (\Nc-2) \over 4} \fSSV (x,x,0)
+ {1\over 4 \Nc} \fSSV (x,x,y) 
\nonumber \\ && 
+ {\Nc-1 \over 4} \fSSV (x,y,x)
\Biggr ]
,\\
V^{(2)}_{VS} &=& g^2 \Bigl [
{\Nc-1 \over 2} \FVS (x,x)
+ {\Nc-1 \over 4} \FVS (x,y)
+ {1\over 2 \Nc} \FVS (y,x)
\nonumber \\ &&
+ {\Nc-1 \over 4 \Nc} \FVS (y,y)
\Bigr ]
,\\
V^{(2)}_{VVS} &=& g^4 \phi^2 \Biggl [ 
{\Nc (\Nc -2 ) \over 2} \FVVS (0,x,0) 
+ {\Nc -1\over 2} \FVVS (x,x,0) 
\nonumber \\ &&
+ {(\Nc-2)^2 \over 2\Nc} \FVVS (x,y,0)
+{(\Nc-1)^2 \over \Nc^2} \FVVS (y,y,0)
\Biggr ] 
,\\
V^{(2)}_{FFV} &=& g^2 \Biggl [ 
{ 2 \Nc^2 - 3 \Nc -1 \over 2} \FFFV (0,x,x) +
\Nc (\Nc -2) \FFFV (x,x,0) 
\nonumber \\ &&
+{\Nc-1 \over 2 \Nc} \FFFV (0,y,y) 
+ {\Nc^2 + 1 \over 2 \Nc} \FFFV (x,x,y) 
+ {3 \Nc - 1 \over 2} \FFFV (x,y,x) 
\Biggr ]
,\\
V^{(2)}_{\FFbV} &=&
-g^4 \phi^2 \Biggl [ 
\Nc (\Nc-2) \FFFbV (x,x,0)
+ \FFFbV (x,x,y)
\nonumber \\ && 
+ 2 (\Nc -1) \FFFbV (x,y,x)
\Biggr ] 
,\\
V^{(2)}_{\rm gauge} &=& g^2 \Biggl [ 
{\Nc (\Nc -2) \over 4} \Fgauge (0,x,x) +
{\Nc \over 4} \Fgauge (x,x,y)
\Biggr ]
.
\eeq
After some algebra, 
using eqs.~(\ref{fSSS})-(\ref{fSSV}) and
(\ref{FVS})-(\ref{Fgauge00x}),
one finds that the sum of these contributions
indeed vanishes, as required by unbroken supersymmetry in the flat
direction
parameterized by $\phi$. 

\subsection{Softly-broken supersymmetric QED}
\label{subsec:softSQED}

Consider the case of supersymmetric QED with a coupling $g$ and
two chiral superfields with charges $\pm1$, as in subsection
\ref{subsec:SQED}. However, now we introduce supersymmetry-breaking
effects in the form of a gaugino mass $M$,
and non-holomorphic soft supersymmetry-breaking scalar squared
masses $m_+^2$ and $m_-^2$ for the scalar fields
of charge $+1$, $-1$ respectively. Instead of equal VEVs,
the scalar fields
of charge $+1$, $-1$ are taken to have classical background values
$\phi$, $0$ respectively.
Then the
four real scalar mass eigenstates obtain squared masses
$x_1,x_1,x_2,x_3$ where
\beq
x_1 &=& m_-^2 - g^2 \phi^2
, \\
x_2 &=& m_+^2 + g^2 \phi^2
, \\
x_3 &=& m_+^2 + 3 g^2 \phi^2, 
\eeq
and the three fermion mass eigenstates obtain squared masses $0,y_1,y_2$,
with
\beq
y_1 &=& [M^2 + 4 g^2 \phi^2 - M \sqrt{M^2 + 8 g^2 \phi^2}]/2
,\\
y_2 &=& [M^2 + 4 g^2 \phi^2 + M \sqrt{M^2 + 8 g^2 \phi^2}]/2
,
\eeq
while the vector
boson obtains a mass
\beq
z = 2 g^2 \phi^2.
\eeq
Because supersymmetry is explicitly
broken, RG invariance requires that a vacuum-energy 
$\Lambda$ is included among the soft supersymmetry breaking terms.
The tree-level potential is then:
\beq
V^{(0)} = \Lambda + m_+^2 \phi^2 + {g^2 \over 2} \phi^4 .
\label{treesoftSQED}
\eeq
From eq.~(\ref{defineoneloop}), the $\overline{\rm DR}'$ one-loop
effective potential 
contribution is:
\beq
V^{(1)} = 2 \oneloop (x_1) + \oneloop (x_2) + \oneloop
(x_3) - 2 \oneloop (y_1) - 2
\oneloop (y_2) + 3 \oneloop (z).
\label{VoneloopsoftSQED}
\eeq
In that scheme, by following the procedures described in sections
\ref{subsec:setup} and \ref{sec:drbarprime}, I
find the following contributions to the
two-loop effective potential:
\beq
V_{SSS}^{(2)} &=& g^4 \phi^2 \Bigl [
\fSSS (x_1,x_1,x_3) 
+ {1\over 2} \fSSS (x_2, x_2, x_3)
+ {3\over 2} \fSSS (x_3, x_3, x_3) \Bigr ] 
,
\label{VsoftSQEDSSS}
\\
V_{SS}^{(2)} &=& g^2 \Bigl [
\fSS (x_1,x_1) 
- {1\over 2} \fSS (x_1, x_2)
- {1\over 2} \fSS (x_1, x_3)
+ {3\over 8} \fSS (x_2,x_2) 
\nonumber \\ &&
+ {1\over 4} \fSS (x_2,x_3) 
+ {3\over 8} \fSS (x_3,x_3) \Bigr ] 
,\\
V_{FFS}^{(2)} &=& {g^2\over M^2 + 8 g^2 \phi^2} \Bigl [
2 (y_1 + z)\fFFS (0,y_1,x_1) 
+2 (y_2 + z)\fFFS (0,y_2,x_1) 
\nonumber \\ &&
+2 z \lbrace \fFFS (y_1,y_1,x_2) 
+\fFFS (y_1,y_1,x_3) 
+\fFFS (y_2,y_2,x_2) 
+\fFFS (y_2,y_2,x_3) \rbrace 
\nonumber \\ &&
+ M^2 \lbrace \fFFS (y_1,y_2, x_2) + \fFFS (y_1,y_2, x_3) \rbrace
\Bigr ] 
, \\
V_{\FFbS}^{(2)} &=& {2 g^4 \phi^2\over M^2 + 8 g^2 \phi^2} 
\Bigl [
2 y_1 \lbrace \fFFbS (y_1,y_1,x_3) - \fFFbS (y_1,y_1,x_2) \rbrace
\nonumber \\ &&
+2 y_2 \lbrace \fFFbS (y_2,y_2,x_3) - \fFFbS (y_2,y_2,x_2) \rbrace
\nonumber \\ &&
+M^2 \lbrace \fFFbS (y_1,y_2,x_2) - \fFFbS (y_1,y_2,x_3) \rbrace
\Bigr ] 
, \\
V_{SSV}^{(2)} &=& {g^2 \over 2} \Bigl [
\fSSV (x_1,x_1,z) + \fSSV (x_2,x_3,z) \Bigr ]
, \\
V_{VS}^{(2)} &=& g^2 \Bigl [
\FVS (z,x_1) 
+ {1\over 2} \FVS (z,x_2) 
+ {1\over 2} \FVS (z,x_3) 
\Bigr ]
, \\
V_{VVS}^{(2)} &=& g^2 z \FVVS(z,z,x_3)
, \\
V_{FFV}^{(2)} &=& {g^2 \over 2(M^2 + 8 g^2 \phi^2)} \Bigl [
(M^2 + 8 g^2 \phi^2) \FFFV (0,0,z)
+y_2 \FFFV (y_1,y_1,z)
\nonumber \\ &&
+y_1 \FFFV (y_2,y_2,z)
+2 z \FFFV (y_1,y_2,z)
\Bigr ]
, \\
V_{\FFbV}^{(2)} &=& {2 g^6 \phi^4 \over M^2 + 8 g^2 \phi^2} 
\Bigl [
\FFFbV (y_1,y_1,z)
+\FFFbV (y_2,y_2,z)
-2 \FFFbV (y_1,y_2,z)
\Bigr ]
, \\
V_{\rm gauge}^{(2)} &=& 0 .
\label{VsoftSQEDgauge}
\eeq

We can now test the RG invariance of the effective
potential. The one-loop scalar anomalous dimension and beta functions in
the $\overline{\rm DR}'$ scheme are:
\beq
\gamma_\phi^{(S,1)} &=& -g^2 
, \label{gamma1softQED}
\\
\beta^{(1)}_g &=& 2 g^3 
, \label{betag1softQED}
\\
\beta^{(1)}_M &=& 4 g^2 M ,\\
\beta^{(1)}_{m^2_+} &=& -8 g^2 M^2 + 2 g^2 (m^2_+ - m^2_-), \\
\beta^{(1)}_{m^2_-} &=& -8 g^2 M^2 + 2 g^2 (m^2_- - m^2_+) ,\\
\beta^{(1)}_{\Lambda} &=& (m^2_+)^2 + (m^2_-)^2 - M^4
.
\label{betaLambda1softQED}
\eeq
From eq.~(\ref{VoneloopsoftSQED}) one therefore finds that
\beq
Q {\partial \over \partial Q} V^{(1)} &=& M^4 - (m^2_+)^2 - (m_-^2)^2 +
8 g^2 \phi^2 M^2 + 2 g^2 \phi^2 m_-^2 
\nonumber \\ && 
- 4 g^2 \phi^2 m_+^2 - 4 g^4 \phi^4,
\eeq
and, from eqs.~(\ref{gamma1softQED})-(\ref{betaLambda1softQED}),
\beq
\sum_I \beta_{\lambda_I}^{(1)} {\partial \over \partial \lambda_I}
V^{(0)} &=& -M^4 + (m^2_+)^2 + (m_-^2)^2 
-8 g^2 \phi^2 M^2 - 2 g^2 \phi^2 m_-^2 
\nonumber \\ && 
+ 2 g^2 \phi^2 m_+^2 + 2 g^4 \phi^4,
\\
- \gamma_\phi^{(S,1)} \phi {\partial \over \partial \phi}
V^{(0)} &=&
2 g^2 \phi^2 m_+^2 + 2 g^4 \phi^4 ,
\eeq
so that eq.~(\ref{oneloopRGcheck}) is satisfied.
At two loop order, one has
\beq
\gamma_\phi^{(S,2)} &=& 4 g^4 ,
\label{gamma2softQED}
\\
\beta^{(2)}_g &=& 8 g^5 ,
\label{betag2softQED}
\\
\beta^{(2)}_M &=& 32 g^4 M
,\\
\beta^{(2)}_{m^2_+} &=& 96 g^4 M^2 + 16 g^4 m^2_+ 
,\\
\beta^{(2)}_{m^2_-} &=& 96 g^4 M^2 + 16 g^4 m^2_- 
,\\
\beta^{(2)}_{\Lambda} &=& 4 g^2 (m^2_+)^2 + 4 g^2 (m^2_-)^2 + 8 g^2 M^4 
,
\label{betaLambda2softQED}
\eeq
so that
\beq
\sum_I \beta_{\lambda_I}^{(2)} {\partial \over \partial \lambda_I}
V^{(0)} &=& 8 g^2 M^4 + 4 g^2 (m^2_+)^2 + 4 g^2 (m_-^2)^2 +
96 g^4 \phi^2 M^2 
\nonumber \\ && 
+ 16 g^4 \phi^2 m_+^2 + 8 g^6 \phi^4,
\label{letsroll}
\\
- \gamma_\phi^{(S,2)} \phi {\partial \over \partial \phi} V^{(0)} &=&
-8 g^4 \phi^2 m_+^2 -8 g^6 \phi^4 .
\eeq
One also finds from eqs.~(\ref{VsoftSQEDSSS})-(\ref{VsoftSQEDgauge}) and
the results of section \ref{sec:drbarprime}:
\beq
Q {\partial \over \partial Q} V^{(2)}  + 
\left [ \sum_I \beta_{\lambda_I}^{(1)} {\partial \over \partial \lambda_I}
- \gamma_\phi^{(S,1)} \phi{\partial \over \partial \phi} \right ]
V^{(1)} &=& 
-8 g^2 M^4 
- 4 g^2 (m_+^2)^2
- 4 g^2 (m_-^2)^2
\nonumber \\
&& - 96 g^4 M^2 \phi^2 - 8 g^4 \phi^2 m_+^2 .
\label{semperfi}
\eeq
Together, eqs.~(\ref{letsroll})-(\ref{semperfi}) verify
eq.~(\ref{twoloopRGcheck}).

\section{Outlook}
\setcounter{equation}{0}
\setcounter{footnote}{1}

In this paper, I have presented the results for the two-loop effective
potential of a general renormalizable field theory in the Landau gauge,
in each of the $\overline{\rm MS}$, $\overline{\rm DR}$, and
$\overline{\rm DR}'$ renormalization schemes. These results should be
useful in connecting specific models of electroweak symmetry breaking to
future data in a precise way. 

It is not unlikely that the
correct model for physics near the TeV scale is based on some 
version of softly-broken
supersymmetry, either the MSSM or some moderate extension of it.
Previous calculations of the effective potential in the MSSM have
used the one-loop result \cite{MSSMonelooppot} and partial two-loop
approximations with leading corrections proportional to $\alpha_S y_t^4$
and $y_t^4$
\cite{Hempfling:1994qq}-\cite{Degrassi:2001yf}. However, there is
still some RG scale-dependence in these results, compared to
estimates of our eventual ability to measure properties of the Higgs
sector at future colliders.
Use of
the
full two-loop
$\overline{\rm DR}'$ effective potential should further reduce the
scale dependence. 
RG
improvement methods 
\cite{Einhorn:1983pp}-%
\cite{Ford:1993mv},
\cite{Yamagishi:1981qq}-\cite{Casas:1999cf}
should enable an accurate determination of the vacuum of the MSSM
and its extensions.
I plan to
report on the application of the results of the present paper to the MSSM
soon \cite{inprogress}.

\addcontentsline{toc}{section}{Appendix A: Individual diagram
contributions to $\fgauge$ and $\Fgauge$}
\section*{Appendix A: Individual diagram 
contributions to the functions $\fgauge$ and
$\Fgauge$}\label{appendix:gauge}
\renewcommand{\theequation}{A.\arabic{equation}}
\setcounter{equation}{0}
\setcounter{footnote}{1}

The three Feynman diagrams labelled $VV$, $VVV$, and $ggV$ in
figure \ref{fig:feynmandiagrams} all involve the field-dependent
coupling $g^{abc}$, and combine to yield $V^{(2)}_{\rm gauge}$.
In the $\overline{\rm MS}$ scheme, the individual diagram contributions to
the function $f_{\rm gauge}(x,y,z)$ are given in an obvious notation by
\beq
\fgauge (x,y,z) &=& 
\fVVV (x,y,z) 
+\fVV (x,y) + \fVV (x,z) + \fVV (y,z)
\nonumber \\ &&
+ \fggV (x) + \fggV (y) + \fggV (z),
\eeq
where
\beq
\fVVV (x,y,z) &=& {1\over 4 x y z} \Bigl \lbrace
(-x^4 - 8 x^3 y - 8 x^3 z +32 x^2 y z +18 y^2 z^2 ) \ihat (x,y,z)
\nonumber \\ &&
+ (y-z)^2 (y^2 + 10 y z + z^2) \ihat (0,y,z) 
- x^4 \ihat (0,0,x)
\nonumber \\ && 
+ (x^2 - 9 y^2 - 9 z^2 + 9 x y + 9 x z -13 yz) x \jhat(y,z)
\nonumber \\ && 
+ 4 x^3 y z + {129\over 4} x y^2 z^2 - 
\left ({20x\over 3} + {y\over 2} + {z\over 2}\right ) xyz J(x) 
\Bigr \rbrace 
\nonumber \\ &&
+ (x \leftrightarrow y) + (x \leftrightarrow z)
, \\
\fVV (x,y) &=& {27 \over 4} \jhat (x,y) 
+ {45 x\over 8} J(y) + {45 y\over 8} J(x)
+ {63 x y \over 16}
, \\
\fggV (x) &=& {x\over 2} \ihat (0,0,x) + {x \over 3} J(x).
\eeq
Similarly, in the $\overline{\rm DR}'$ scheme, 
\beq
\Fgauge (x,y,z) &=& 
\FVVV (x,y,z) 
+ \FVV (x,y) + \FVV (x,z) + \FVV (y,z)
\nonumber \\ &&
+ \fggV (x) + \fggV (y) + \fggV (z),
\eeq
where
\beq
\FVVV (x,y,z) &=& \fVVV (x,y,z) 
+ \feeV (x,y,z)
+ \feeV (z,x,y)
+ \feeV (y,z,x) ,
\\
\FVV (x,y) &=& \fVV (x,y) 
+ \feV (x,y)
+ \feV (y,x)
+ \fee (x,y) .
\eeq
with $\fggV (x)$  given as before. 
Explicitly,
\beq
\FVVV (x,y,z) &=& {1\over 4 x y z} \Bigl \lbrace
(-x^4 - 8 x^3 y - 8 x^3 z 
+32 x^2 y z
+18 y^2 z^2 
) \ihat (x,y,z)
\nonumber \\ &&
+ (y-z)^2 (y^2 + 10 y z + z^2) \ihat (0,y,z) 
- x^4 \ihat (0,0,x)
\nonumber \\ && 
+ (x^2 - 9 y^2 - 9 z^2 + 9 x y + 9 x z -13 yz) x \jhat(y,z)
\nonumber \\ && 
+ {x y^2 z^2 \over 4} 
+(-{44x\over 3}+{47 y\over 2}+{47 z\over 2}) x y z J(x)
\Bigr \rbrace 
\nonumber \\ &&
+ (x \leftrightarrow y) + (x \leftrightarrow z)
, \\
\FVV (x,y) &=& {27 \over 4} \jhat (x,y) 
- {3 x\over 8} J(y) - {3 y\over 8} J(x)
- {x y \over 16}
.
\eeq
The results for vanishing arguments are easily obtained from 
eqs.~(\ref{deltaforce})-(\ref{sealteamsix}).

\addcontentsline{toc}{section}{Appendix B: Renormalization-group-scale
derivatives}
\section*{Appendix B: Renormalization-group-scale
derivatives}\label{appendix:Qderivs}
\renewcommand{\theequation}{B.\arabic{equation}}
\setcounter{equation}{0}
\setcounter{footnote}{1}

It is often useful to have expressions for the derivatives of the
two-loop effective potential functions with respect to the
renormalization scale $Q$, for example to check RG invariance. The
derivative of the one-loop effective potential
function $\oneloop (x)$ defined in eq.~(\ref{defineoneloop}) is 
\beq
Q{\partial \over \partial Q} \oneloop (x) = -x^2/2 
. 
\eeq
For checking the RG-invariance of the effective potential, it is 
convenient to write the derivatives of two-loop functions with respect to
$Q$ in terms of the derivative of the one-loop function with respect
to its squared-mass argument:
\beq
h'(x) = {x\over 2} (\lnbar x-1) .
\eeq
The derivatives of the two-loop functions can all be found
from those of the basis functions:
\beq
Q{\partial \over \partial Q} J(x) &=& -2 x
, \\
Q{\partial \over \partial Q} \jhat (x,y) &=& 
-4 y h'(x) -4 x h'(y)
, \\
Q{\partial \over \partial Q} \ihat (x,y,z) &=&
4[ h'(x) + h'(y) + h'(z) ] - 2(x+y+z)
.
\eeq
For the derivatives of the $\overline{\rm MS}$ two-loop functions,
one finds:
\beq
Q{\partial \over \partial Q} \fSSS (x,y,z) &=&
-4[ h'(x) + h'(y) + h'(z) ] + 2(x+y+z)
,
\label{dQfsss}
\\
Q{\partial \over \partial Q} \fSS (x,y) &=&
-4 y h'(x) -4x h'(y)
,
\label{dQfss}
\\
Q{\partial \over \partial Q} \fFFS (x,y,z) &=&
4 x h'(x) + 4 y h'(y) + (8 x +8y - 4 z) h'(z)
\nonumber \\ && 
-2 x^2 - 2y^2 + 2 z^2 - 4 xy
, \\
Q{\partial \over \partial Q} \fFFbS (x,y,z) &=&
8 [ h'(x) + h'(y) + h'(z)] - 4 (x+y+z)
,
\label{dQfffbs}
\\
Q{\partial \over \partial Q} \fSSV (x,y,z) &=&
12 x h'(x) + 12 y h'(y) + (12 x + 12 y -4 z) h'(z) 
\nonumber \\ && 
-2 x^2 - 2y^2 -12 xy -6 xz -6 yz + {10 \over 3} z^2 
,
\\
Q{\partial \over \partial Q} \fVS (x,y) &=& 
-12 y h'(x) - 12 x h'(y) -4 xy
, \\
Q{\partial \over \partial Q} \fVVS (x,y,z) &=&
-9 h'(x) - 9 h'(y) -12 h'(z) + {9 x\over 2} + {9 y \over 2} -z
, \\
Q{\partial \over \partial Q} \fFFV (x,y,z) &=&
(12 x + 12 y - 8z) h'(z) + 6 xz + 6 yz + {8 z^2 \over 3}
, \\
Q{\partial \over \partial Q} \fFFbV (x,y,z) &=&
24 [h'(x) + h'(y) + h'(z)] - 4x -4y-12z
, \\
Q{\partial \over \partial Q} \fgauge (x,y,z) &=& 
9 [ (x+y) h'(z) + (x+z) h'(y) + (y+z) h'(x)]
\nonumber \\ &&
+ 52 [ x h'(x) + y h'(y) + z h'(z) ]
\nonumber \\ &&
-{19 \over 3} (x^2 + y^2 + z^2)
-63 (xy + yz + xz)
.
\eeq
For the functions used with epsilon scalars in the $\overline{\rm DR}$
scheme,
one has:
\beq
Q{\partial \over \partial Q} \feS (x,y) &=& 4xy
, \\
Q{\partial \over \partial Q} \feeS (x,y,z) &=& 4z
, \\
Q{\partial \over \partial Q} \fFFe (x,y,z) &=& -4 x^2 -4 y^2
, \\
Q{\partial \over \partial Q} \fFFbe (x,y,z) &=& -8x-8y
, \\
Q{\partial \over \partial Q} \feV (x,y) &=& 12 xy
, \\
Q{\partial \over \partial Q} \fee (x,y) &=& 0
, \\
Q{\partial \over \partial Q} \feeV (x,y,z) &=& -12 xz -12 yz + 4 z^2 
. \eeq
Finally, the functions used in the $\overline{\rm DR}'$ scheme 
(besides those found in $\overline{\rm MS}$) satisfy:
\beq
Q{\partial \over \partial Q} \FVS (x,y) &=& 
-12 y h'(x) - 12 x h'(y)
, \\
Q{\partial \over \partial Q} \FVVS (x,y,z) &=&
-9 h'(x) - 9 h'(y) -12 h'(z) + {9 x\over 2} + {9 y \over 2} +3z
, \\
Q{\partial \over \partial Q} \FFFV (x,y,z) &=&
(12 x + 12 y - 8z) h'(z) -4 x^2 - 4y^2 + 6 xz + 6 yz + {8 z^2 \over 3}
, \\
Q{\partial \over \partial Q} \FFFbV (x,y,z) &=&
24 [h'(x) + h'(y) + h'(z)] - 12(x+y+z)
, \\
Q{\partial \over \partial Q} \Fgauge (x,y,z) &=& 
9 [ (x+y) h'(z) + (x+z) h'(y) + (y+z) h'(x)]
\nonumber \\ &&
+ 52 [ x h'(x) + y h'(y) + z h'(z) ]
\nonumber \\ &&
-{7 \over 3} (x^2 + y^2 + z^2)
-63 (xy + yz + xz)
.
\eeq

\bigskip \noindent {\it Acknowledgements:}   This work was
supported in part by the National Science Foundation
grant number PHY-9970691.


\end{document}